\documentclass[12pt,preprint,usenatbib]{emulateapj}

\usepackage{amssymb}
\usepackage{times}
\usepackage{pifont} 
\usepackage{rotating}

\newcommand{\muhz}{$\mu$Hz}
\newcommand{\numax}{$\nu_{\mathrm{max}}$}
\newcommand{\dnu}{$\Delta\nu$}
\newcommand{\astec}{\textsc{ASTEC}}
\newcommand{\mesa}{\textsc{MESA}}
\newcommand{\adipls}{\textsc{ADIPLS}}

\newcommand{\kepler}{\textit{Kepler}}
\newcommand{\corot}{\textit{CoRoT}}

\newcommand{\dP}{$\Delta P$}
\newcommand{\dPa}{$\Delta P_\mathrm{as}$}

\newcommand{\msol}{M$_\odot$}

\def\Mo{\rm{M}_\odot}
\def\muHz{\mu{\rm Hz}}
\def\note #1]{\noindent{\bf #1]}}

\def\sgnk2{{\rm sgn\left(K^2\right)}}
\def\rmd{{\rm d}}
\def\rmdd{{\rm d}^2}
\def\mN{\mathcal{N} }
\def\mH{\mathcal{H} }
\def\mF{\mathcal{F} }
\def\mG{\mathcal{G} }

\def\brunt{buoyancy frequency }
\def\numax{$\nu_{\rm max}$}
\def\brunts{buoyancy frequencies}

\shorttitle{Buoyancy glitches in the cores of red giants}
\shortauthors{Cunha et al.}

\begin{document}

\title{Structural glitches near the cores of red giants revealed by
  oscillations in g-mode period spacings from stellar models}
\author{M.~S.~Cunha~\altaffilmark{1}, D. Stello~\altaffilmark{2,3},
  P.P. Avelino~\altaffilmark{1,4} ,
  J. Christensen-Dalsgaard~\altaffilmark{3} , R.~H.~D. Townsend~\altaffilmark{5}
}
\altaffiltext{1}{Instituto de Astrof\'{\i}sica e Ci\^encias do Espa\c co,
  Universidade do Porto, CAUP, Rua das Estrelas, 4150-762 Porto, Portugal, mcunha@astro.up.pt}
\altaffiltext{2}{Sydney Institute for Astronomy (SIfA), School of Physics, University of Sydney, NSW 2006, Australia}
\altaffiltext{3}{Stellar Astrophysics Centre, Department of Physics
  and Astronomy, Aarhus University, Ny Munkegade 120, DK-8000 Aarhus
  C, Denmark}
\altaffiltext{4}{Departamento de F\'{\i}sica e Astronomia, Faculdade de Ci\^encias, Universidade do Porto, Rua do Campo Alegre 687, PT4169-007 Porto, Portugal}
\altaffiltext{5}{Department of Astronomy, University of Wisconsin-Madison, 2535 Sterling Hall, 475 N. Charter Street, Madison, WI 53706, USA}

\begin{abstract}
With recent advances in asteroseismology it is now possible to peer into the cores of red giants, potentially providing a way to study processes such as nuclear burning and mixing through their imprint as sharp structural variations -- glitches -- in the stellar cores. Here we show how such core glitches can affect the oscillations we observe in red giants.  We derive an analytical expression describing the expected frequency pattern in the presence of a glitch.  This formulation also accounts for the coupling between acoustic and gravity waves.  From an extensive set of canonical stellar models we find glitch-induced variation in the period spacing and inertia of non-radial modes during several phases of red-giant evolution.  Significant changes are seen in the appearance of mode amplitude and frequency patterns in asteroseismic diagrams such as the power spectrum and the \'echelle diagram. Interestingly, along the red-giant branch glitch-induced variation occurs only at the luminosity bump, potentially providing a direct seismic indicator of stars in that particular evolution stage.  Similarly, we find the variation at only certain post-helium-ignition evolution stages, namely, in the early phases of helium-core burning and at the beginning of helium-shell burning signifying the asymptotic-giant-branch bump. Based on our results, we note that assuming stars to be glitch-free, while they are not, can result in an incorrect estimate of the period spacing. We further note that including diffusion and mixing beyond classical Schwarzschild, could affect the characteristics of the glitches, potentially providing a way to study these physical processes.

\end{abstract}

\keywords{stars: evolution --- stars: oscillations --- stars:
  interiors}


\section{Introduction} 

The cores of red-giant stars hold the key to answering a number of unresolved
questions about fundamental physics that govern stellar evolution, 
such as mixing process, rotation, and the effect of magnetic fields. 
It has been known for over a decade that red giants show
stochastically-driven oscillations like the Sun \citep{frandsen02}, but
only with recent data from space missions like \corot\ and \kepler, have
asteroseismic investigations revealed details about the cores of red giants.
This advance has been possible due to the fortunate circumstance that 
gravity waves (hereafter, g modes) in the cores of red giants couple to 
acoustic waves (hereafter, p modes) in the envelope, resulting in mixed
modes whose information about the core properties is therefore observable
at the surface~\citep{dupret09,bedding10}. 
Recent findings include the distinction between stars with inert cores from 
those that possess core burning \citep[e.g.~][]{bedding11}, the measurement
of core rotation rates much slower than predicted by current
theory of angular momentum transport
\citep[e.g.~][]{beck12,mosser12c,cantiello14}, and an ability to determine
the evolutionary stages of stars with unprecedented precision  
\citep{mosser14}. 
Despite these findings, the full potential of current asteroseismic
data can only be realized if all aspects are understood about how the
internal structure of stars may influence the observed oscillations. To achieve 
this, it is necessary to explore how sharp structural variations inside a red giant
could impact its oscillation frequencies. Sharp structural variations can 
be found in stellar interiors at the borders of convectively mixed 
regions, in regions of ionization of elements, or between
layers that have acquired different chemical composition as a result of
nuclear burning.  The signatures they imprint on the oscillation frequencies
have already been studied observationally and theoretically in white dwarfs
\citep[e.g.][]{winget91,brassard92}, main-sequence stars
\citep{roxburgh01,miglio08,degroote10} including those like the Sun
\citep{monteiroetal00,monteiro05,cunha07,cunha11,Houdek07,mazumdar14}, and in sdB stars
\citep{charpinet00,ostensen14}, which are essentially the
cores of previous red giants.   

In red giant stars, only the signature of the helium ionization zone has
been studied \citep{miglio10}. This signature arises from the stellar
envelope and affects the acoustic modes, but simulations indicate its
application as a diagnostic tool on single stars might be limited
\citep{broomhall14}. 
However, the effect of sharp variations occurring in the deeper layers near the
cores of red giants has neither been investigated theoretically, nor been
reported from observations.  Given the mixed character of the waves in red
giants, the study of this phenomenon requires understanding the 
combined effect of the sharp structural variation and of the coupling between
p and g modes.

Here we present the first comprehensive study of the effect on the
oscillation frequencies of red giants from sharp structural variations
located in their deeper layers.  We illustrate the impact this may have 
on common asteroseismic diagrams and investigate where this effect might be relevant during the red
giant evolution phase. While no assumption is made about the degree of
the modes in the analysis presented here, all examples
provided are for dipole modes, because these are the most promising
from the observational point of view.

      \begin{figure} 
\centering
\hspace{-0.5cm}
               \rotatebox{0}{\includegraphics[width=8.5cm]{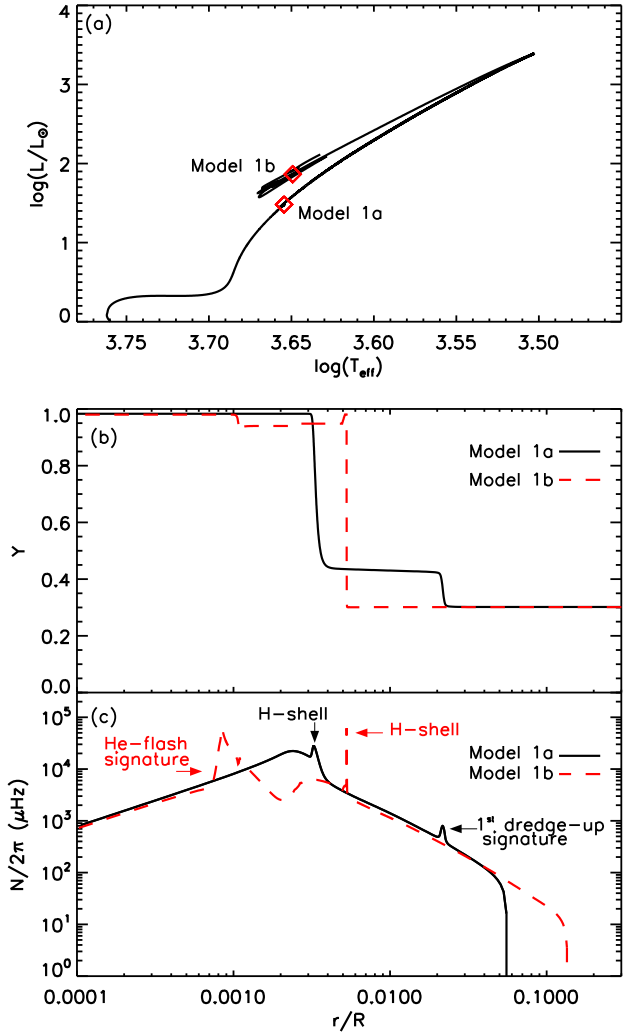}}
  \caption{One solar mass red-giant models considered for a detailed
       analysis. {\bf (a)} Position in the HR diagram: model~1a is on
       the  red-giant branch just below the luminosity bump and
       model~1b is between helium flashes.; {\bf (b)} and {\bf (c)}
       show, respectively, the helium profile and 
       buoyancy frequency in the inner region of model~1a (solid curve)
       and model~1b (dashed curve). The sudden decrease of the buoyancy
       frequency, at $r/R\sim 0.055$ for model~1a and $r/R\sim 0.135$
       for model~1b, marks the lower boundary of the convective envelope in
       the corresponding model.
  }
     \label{Brunt_both}
   \end{figure}

\section{Structure of the g-mode cavity}
\label{structure}

Internal gravity waves have frequencies below the
buoyancy (or Brunt-V\"ais\"al\"a) frequency and propagate only where
there is no convection. While on the red-giant branch a star is
powered by hydrogen burning in a shell
surrounding an innert radiative helium core. The g-mode
propagation cavity extends essentially from the stellar center to the bottom of
the convective envelope. Once stable core-helium burning starts, the
central part of the core becomes convective, reducing the size of the g-mode
cavity.  For massive stars the transition
between these two phases is smooth. However, according to current
standard 1D stellar models, in lower-mass stars with a
degenerate helium core, this transition involves a succession of
off-centered helium flashes \citep{bildsten12} \cite[see also][ for a general overview of red-giant evolution]{salaris02}.

The propagation speed of the gravity
waves depends on the buoyancy frequency. Consequently, variations in the buoyancy
frequency inside the g-mode cavity may perturb the periods of high-radial-order modes
away from their asymptotic value. 
Sharp variations in the \brunt during the red-giant phase usually result from local changes
in the chemical composition. Examples of these variations are illustrated in
Figure~\ref{Brunt_both} where we show two red-giant models at
different evolution stages (panel a), prior to and during the
helium-flash phase, respectively, and their corresponding helium
abundances  (panel b) and buoyancy frequencies, $N$ (panel c), for the core
region, where $N$ is defined by the relation,
\begin{eqnarray}
N^2=g\left(\frac{1}{\gamma_1}\frac{\rmd\ln p}{\rmd r}-\frac{\rmd \ln\rho}{\rmd r}\right).
\label{bruntdef}
\end{eqnarray}
Here, $r$ is the distance from the stellar center
in a spherical coordinate system ($r$,$\theta$,$\varphi$) and $g$,
$\gamma_1$, $p$ and $\rho$ are, respectively,  the gravitational
acceleration, the first adiabatic exponent, the pressure, and the density
in the model.
The models were computed with the evolution codes
\astec~\citep{jcd08a} and \mesa~\citep{paxton13}, respectively. 
Two
spikes are visible in the \brunts. The spikes located at relative radii of $\approx 0.003$
(model~1a) and $\approx 0.005$ (model~1b)
result from the chemical-composition variation at the 
hydrogen-burning shell. The spike furthest out
in model~1a, at a 
relative radius of $\approx 0.02$,  results from strong chemical
gradients left behind by the retreating
convective envelope which, during the first dredge-up, extended to the
region where the gas had previously been processed by nuclear
burning.\footnote{Since the model does not include diffusion, the
  dredge-up should leave behind a discontinuity in
  composition. However, the numerical treatment of the mesh in the
  ASTEC calculation causes numerical diffusion which leads to some
  smoothing of the composition profile and hence broadening and
  lowering of the buoyancy-frequency spike, as is evident in Fig. 1. A
  similar but less pronounced effect appears to be present in the MESA models.} As the convective envelope retreats, the g-mode cavity expands to 
include the sharp variation in the chemical composition; 
this eventually disappears, when reached 
by the hydrogen-burning shell which is moving out in mass as the helium core grows. 
In the case of low-mass stars, this takes place 
while the star is still on its way up the red-giant branch, 
when it reaches the well-known luminosity bump. 
The bump shows itself as a temporary decrease in luminosity 
when the hydrogen-burning shell gets close to the sharp variation in the chemical composition. As a result of the decrease 
in the average mean molecular weight in the region just above the
shell,  the luminosity of the hydrogen-burning shell decreases. This
is  followed by a return to increasing luminosity when the hydrogen-burning shell reaches the sharp variation.\footnote{For stars more massive than 2.2 $\Mo$ helium burning is ignited before the 
  hydrogen-burning shell reaches the discontinuity and no bump occurs on the red-giant branch.} ~(Hekker and Christensen-Dalsgaard, in preparation).  Finally, the
innermost
spike in model~1b, at a relative radius of $\approx 0.0008$, results from
the chemical composition variation caused by a helium flash. Spikes in
the \brunt may have yet a different origin from those discussed
above. In particular, they can result from sharp variations in
chemical composition left by retreating convective cores that were
active either during the main sequence or during the
helium-core-burning phase. These will be illustrated in
section~\ref{evolution} where we look at sharp buoyancy variations
along the red-giant evolution more broadly.

\begin{figure*} 
          \begin{minipage}{0.44\linewidth}
         \hspace{0.04\linewidth}
             \rotatebox{270}{\includegraphics[width=0.55\linewidth]{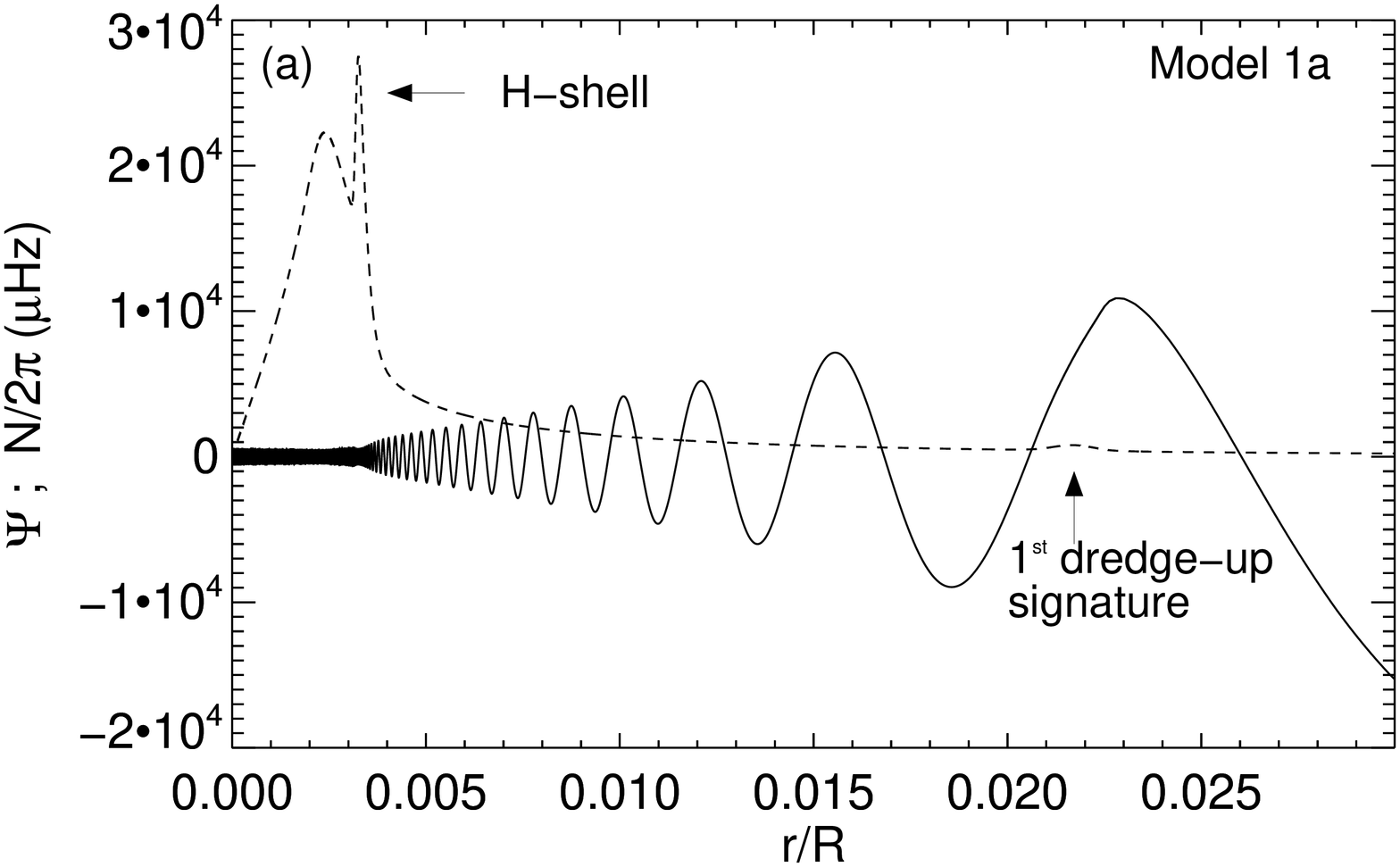}}
        \end{minipage}
           \hspace{0.08\linewidth}
           \begin{minipage}{0.44\linewidth}
             \hspace{-0.01\linewidth}
              \rotatebox{270}{\includegraphics[width=0.55\linewidth]{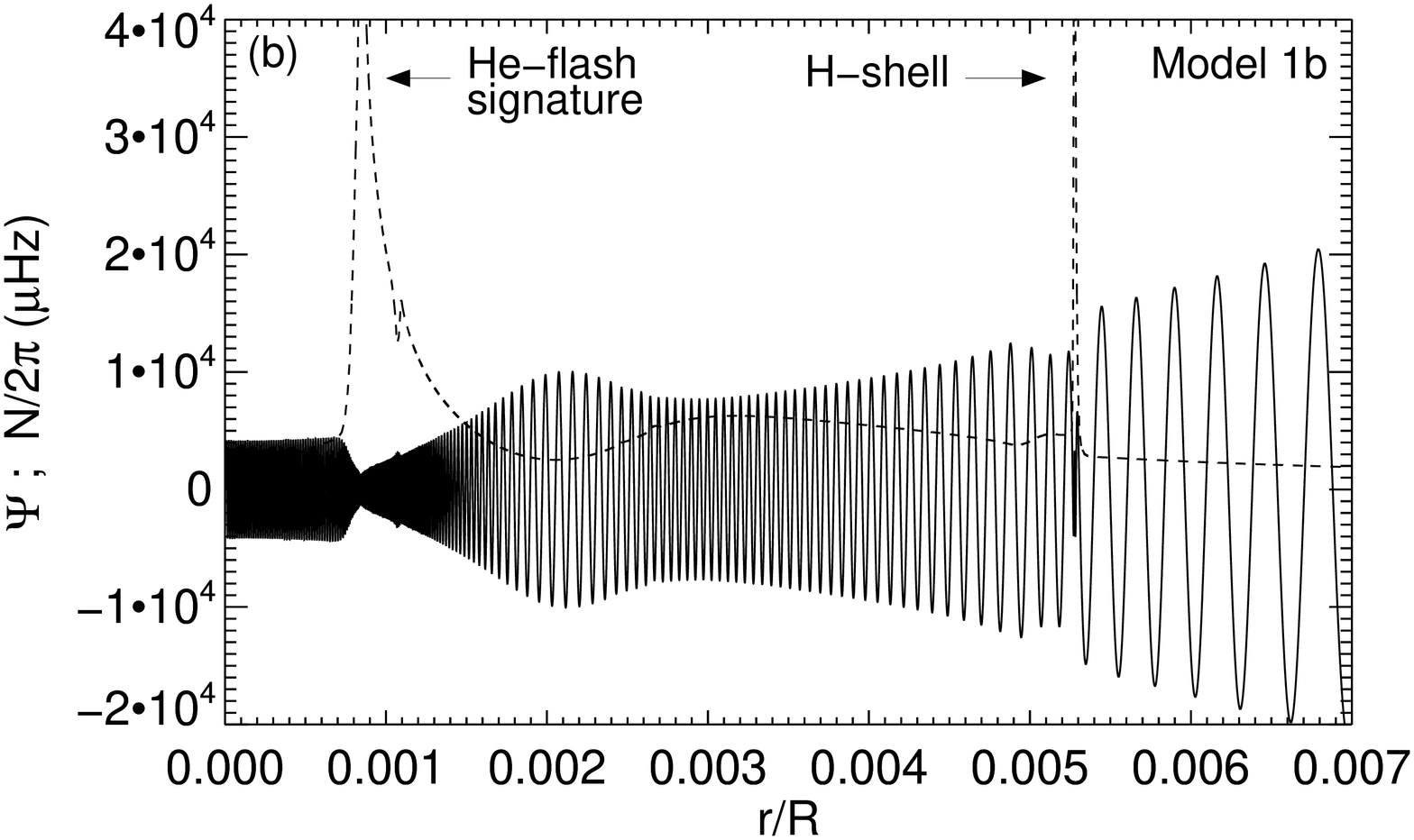}}
        \end{minipage}
\vspace{1.cm}
          \caption{Asymptotic eigenfunction
            (solid curve) and the buoyancy
            frequency (dashed curve) for:   {\bf (a)} model~1a  and
            {\bf (b)} model~1b.
            The eigenfunctions have arbitrary
            amplitude and are for characteristic 
            eigenfrequencies of these models. The arrows mark the
            positions of the \brunt spikes discussed in the text and
            seen also in Figure~\ref{Brunt_both}c.  }
         \label{figeigen}

          \end{figure*}

Whether or not the spikes in the buoyancy frequency are sufficiently
sharp to produce a significant deviation of the frequencies of high-radial-order g modes
from their asymptotic value depends on how the characteristic width of the spikes compares with
the local wavelength. A comparison of the two scales is illustrated in
Figure~\ref{figeigen} for the two models presented in
Figure~\ref{Brunt_both}. The eigenfunction $\Psi$ shown in this
figure is related to the Lagrangian pressure perturbation
(see section~\ref{g-modes} for a precise definition).  Clearly, the
width of the inner spike
 is much larger than the local wavelength in both models. Hence, this
 spike is seen as a smooth variation by
the wave and is well accommodated by asymptotic analysis. In
contrast, at the outermost spikes the buoyancy frequency varies at a scale
comparable to or shorter than the local wavelength. We therefore may
expect these features - {\it hereafter glitches} - to change the oscillation frequencies
 from their asymptotic value. In that case, the period spacing may also deviate from the fixed value
predicted by the asymptotic theory \citep{tassoul80}. 

\section{Glitch effect on the period spacing: toy model}
\label{g-modes}

In this section we illustrate the effect of a buoyancy glitch on the oscillation frequencies and, consequently, on
the period spacing.  To accomplish that we consider first an
analytical toy model in
which the glitch is assumed to be infinitely narrow and well modeled by
a Dirac delta function. In the analysis we first introduce the
analytical description of the problem, then consider the effect of the glitch on
pure g modes and, finally, consider the same effect when the latter
couple to the envelope p modes. 

\subsection{Setting the problem}
\label{analytical}

Our starting point for the analytic analysis is a second-order
differential equation for the radial dependent part of the Lagrangian pressure perturbation,
$\delta p$, derived from the equations that describe linear, adiabatic
perturbations to a spherically symmetric star, under the Cowling
approximation (hence neglecting the Eulerian perturbation to the
  gravitational potential). This equation can be written in the
  standard wave-equation form \citep{gough93,gough07} by adopting $\Psi=~(r^3/g\rho f) ^{1/2}\delta p$ as the
dependent variable, where $f$ is a function of frequency and of the
equilibrium structure (the f-mode discriminant defined by equation
(35) of \cite{gough07}). In terms of this variable, the wave equation takes
the form,
\begin{eqnarray}
\frac{\rmdd\Psi}{\rmd r^2}+K^2\Psi=0,
\label{waveeqap}
\end{eqnarray}
with the radial wavenumber $K$ defined by,
\begin{eqnarray}
K^2=\frac{\omega^2-\omega_{\rm
    c}^2}{c^2}-\frac{L^2}{r^2}\left(1-\frac{\mN^2}{\omega^2}\right).
\label{k2}
\end{eqnarray}
Here, $L^2=l(l+1)$ and $l$ is the angular degree of the mode, $c$ is the sound speed, and $\omega_{\rm c}$ and $\mN$ are
generalizations of the usual critical acoustic frequency and
buoyancy frequency, respectively, which account for all
terms resulting from the spherical geometry of the problem.  The exact forms of
these quantities can be found in equations (5.4.8) and (5.4.9) of
\cite{gough93}, which are reproduced in Appendix B of this paper.  The radii where $K^2 = 0$ define the turning points of
the modes. These separate the regions where waves can propagate (where
$K^2 >0$) from where they are evanescent (where $K^2
< 0$). 

   \begin{figure}
 \centering
 \rotatebox{270}{\includegraphics[width=0.7\linewidth]{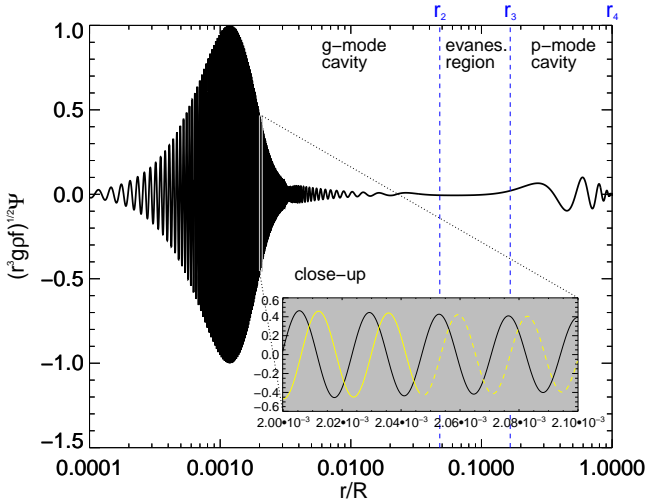}}
\vspace{1.cm}
     \caption{Normalized eigenfunction, as function of relative
       radius, for the dipole mode with frequency  $\nu = 51.20$~$\mu$Hz, computed with the pulsation
       code \adipls\ for our model~1a. The chosen eigenfunction  $(r^3g\rho
       f) ^{1/2}\Psi=~r^3\delta p$ has the dimensions of energy and
       is normalized to be 1 at its maximum value.  The vertical, blue
       dashed lines show the position of $r_2$ and $r_3$, the  two turning
       points bounding the evanescent region. The outermost turning
       point, $r_4$, is also shown, while the innermost turning point, $r_1$, is
       outside the plotted range. The
       g-mode cavity is located between the unseen $r_1$ and $r_2$ and the p-mode cavity is
       located between $r_3$ and $r_4$.  The close-up shows a comparison
       between the numerical (in black) and analytical (in yellow) eigenfunctions in a particular region, well
       inside the g-mode cavity. The continuous yellow curve
       represents the inner solution derived from
       equation~(\ref{gasymp_l}), while the dashed yellow curve
       represents the outer solution derived from
       equation~(\ref{gasymp_r}). 
  }
     \label{twowaves}
   \end{figure}

For typical red giants, including the models discussed in
section~\ref{structure}, there are two separate propagation regions
defined by four turning points. We denote these points as $r_1$,
$r_2$, $r_3$, and $r_4$.  We note that for the
models under consideration, $r_1$ and $r_4$ are essentially at the center of the
star ($r=0$) and at the stellar photosphere ($r=R$), respectively.  
The propagation regions and the turning points $r_2$, $r_3$, and $r_4$ are illustrated in
Figure~\ref{twowaves} for a representative dipole mode 
in our model~1a, where we show the corresponding mode
eigenfunction derived with the pulsation code \adipls\
\citep{jcd08b}.  The evanescent region, which is located between the turning points
$r_2$ and $r_3$, separates the two
propagation cavities. To its left is the so-called g-mode 
cavity while to its right we have the p-mode cavity. 

In practice, $N$ (equation~(\ref{bruntdef})) is a very good approximation of $\mN$ everywhere 
except very close to the center of the star and in the evanescent
region between the two cavities, where the latter diverges. These
differences will be fully accounted for in future work
(Cunha et al., in preparation).  Nevertheless, in the toy
model presented here, we will approximate $\mN$ by $N$ from the outset. Despite this and other approximations
that will follow, our toy model retains all important features seen in the full numerical
solutions obtained with \adipls\ and, as will become clear in
section~\ref{fullnumerical},  will be important for the
correct interpretation of the results of the latter. 

\subsection{Effect on pure g modes}
\label{ganalytical}

\subsubsection{The eigenvalue condition}
\label{eigencond}
To understand the impact of a glitch on the oscillation
frequencies it is convenient to start by analyzing a simpler problem in
which we ignore any coupling between the g and p modes. This coupling will be considered in section~\ref{coupling}.

 In order to find the oscillation frequencies for the pure
 g modes we need to impose adequate boundary conditions to the
 solution of equation~(\ref{waveeqap}). Towards the center of the star
 this condition is that $\Psi$ decreases exponentially as $r$ goes to
 zero. Moreover, because we are ignoring any coupling and because the g-mode cavity is located at such depth that the
 stellar atmosphere hardly influences the solutions, 
 the condition towards the envelope also needs to be that $\Psi$ decreases exponentially for $r
 \gg r_2$. From the asymptotic analysis of equation~(\ref{waveeqap}),
 which ignores the effect of the glitch, we know that the
 solution, $\Psi_{\rm in}$, satisfying the inner boundary condition has
the form \citep{gough93}, 
\begin{equation}
\Psi_{\rm in} \sim \tilde\Psi_{\rm in} K_0^{-1/2} \sin\left(\int_{r_1}^r K_0\rmd r +
  \frac{\pi}{4}\right),
\label{gasymp_l}
\end{equation}
 in the region $ r_1 \ll r \ll r_2$, where, following the notation of
 Gough, we have used the symbol $\sim$ to indicate that the two sides of the
 equation are asymptotically equal. Here, $\tilde\Psi_{\rm in}$ is a constant and
 the subscript {\small 0} on $K$ indicates that we are not accounting
 for the glitch. 
 This inner solution is illustrated in the inset in Figure~\ref{twowaves} by the continuous yellow curve. 
Likewise, the asymptotic solution to equation~(\ref{waveeqap}), $\Psi_{\rm out}$, 
that satisfies the outer boundary condition can be written as \citep{gough93},
\begin{equation}
\Psi_{\rm out} \sim \tilde\Psi_{\rm out} K_0^{-1/2} \sin\left(\int_r^{r_2} K_0\rmd r +
  \frac{\pi}{4}\right),
\label{gasymp_r}
\end{equation}
in the region $ r_1 \ll r \ll r_2$,
where $\tilde\Psi_{\rm out}$ is also a constant. This outer solution is illustrated in
Figure~\ref{twowaves} (inset)  by the dashed yellow curve. 

Since equations~(\ref{gasymp_l}) and (\ref{gasymp_r}) are both
valid well inside the g-mode cavity, they must be the
same. The requirement that they be the same provides the
eigenvalue condition (the condition that determines which
oscillation (eigen)frequencies are allowed by the above boundary conditions). In this case, the eigenvalue condition translates to
\begin{equation}
\int_{r_1}^{r_2}K_0\rmd r = \pi\left(n-\frac{1}{2}\right),
\label{eigen}
\end{equation}
where $n$ is a positive integer. Hence, it is this condition that
ensures the two yellow curves match (Figure~\ref{twowaves} (inset)).
The phase shift that these solutions show in relation to the full
\adipls\ solution (solid, black curve) is due to their not including the coupling to the
p modes.

Next, we include the effect from a glitch in the buoyancy frequency. To
keep the toy model simple we will initially assume
that the glitch appears at a single position in
radius, $r=r_\star$, well inside the g-mode cavity, such that the asymptotic solutions (\ref{gasymp_l}) and (\ref{gasymp_r}) are
still valid on either side of it (this assumption will be relaxed in
section~\ref{gnumerical}). 
Accordingly, we represent the glitch by a Dirac delta function,
$\delta$, such that the
buoyancy frequency becomes,
\begin{eqnarray}
{N^2}={N_0^2}\left[1+A\delta\left(r-r_\star\right)\right],
\label{glitch}
\end{eqnarray}
where  $A$  has dimensions of length and is a measure of the strength of the glitch, and $N_{0}$ is
the glitch-free buoyancy frequency.
By imposing continuity of the solutions~\footnote{Strictly speaking,
  the continuity condition is satisfied by $\delta p$.  However, we
  have verified from the numerical solutions computed with \adipls\ that this condition is
  also very closely satisfied by $\Psi$.} given by equations~(\ref{gasymp_l}) and (\ref{gasymp_r})  at $r=r_\star$ we find,
\begin{eqnarray}
\tilde\Psi_{{\rm in}}=\frac{\sin\left(\int_{r_\star}^{r_2} K_0\rmd r +
  \frac{\pi}{4}\right)}{\sin\left(\int_{r_1}^{r_\star} K_0\rmd r +
  \frac{\pi}{4}\right)}\tilde\Psi_{{\rm out}}.
\label{amplcond}
\end{eqnarray}

Because under the approximation considered here the glitch is infinitely narrow, the first derivative of the solution is not
continuous at $r=r_\star$. The condition to be imposed on the
derivative can be found by
integrating the wave equation (\ref{waveeqap}) once in a finite
region of width $2\epsilon$ across the glitch and then taking the
limit when $\epsilon$ goes to zero.  Accordingly, we have,
\begin{eqnarray}
\int_{r_\star-\epsilon}^{r_\star+\epsilon}\frac{\rmdd\Psi}{\rmd r^2}+\int_{r_\star-\epsilon}^{r_\star+\epsilon} K^2\Psi=0,
\label{waveeqint}
\end{eqnarray}
where now $K$ takes the glitch into account, differing from $K_0$ only
at $r=r_\star$,  where $N$ differs from $N_0$. 
Well inside the g-mode cavity $K$ (equation~(\ref{k2}), with $\mathcal
N$
replaced by $N$ ) may be 
approximated by,
\begin{eqnarray}
K\approx\frac{L N}{\omega \, r},
\label{kg}
\end{eqnarray}
and, thus,  we  find, 
\begin{eqnarray}
\left|\frac{\rmd\Psi_{\rm out}}{\rmd r}-\frac{\rmd\Psi_{\rm in}}{\rmd
  r}\right|_{r_\star}=-A K_0^2\left(r_\star\right)\Psi\left(r_\star\right),
\label{derivcond}
\end{eqnarray}
when $\epsilon\to 0$.

By differentiating equations~(\ref{gasymp_l}) and (\ref{gasymp_r}) and neglecting
the small terms resulting from the derivatives of the amplitudes, $K_0^{-1/2}$, we
find,  after substituting in  equation~(\ref{derivcond}),
\small
\begin{eqnarray}
\tilde\Psi_{\rm out}K_0^{1/2} \left(r_\star\right)
\cos\left(\int_{r_\star}^{r_2}K_0\rmd r+ \frac{\pi}{4}
\right) + & & \nonumber \\ \tilde\Psi_{\rm in}K_0^{1/2}\left(r_\star\right)\cos\left(\int_{r_1}^{r_\star}K_0\rmd
  r+\frac{\pi}{4}\right) = & & \nonumber \\  && \hspace{-4cm}  A \tilde\Psi_{\rm out}K_0^{3/2}\left(r_\star\right)\sin\left(\int_{r_\star}^{r_2}K_0\rmd
  r+\frac{\pi}{4} \right). 
\label{eigengap_int} 
\end{eqnarray}
\normalsize
Using the continuity condition (\ref{amplcond}) and the fact that
$K_0(r_\star)\approx LN_0(r_\star)/\omega r_\star\equiv LN_0^\star/\omega
\, r_\star$, equation~(\ref{eigengap_int}) becomes
\small
\begin{eqnarray}
\sin\left(\int_{r_1}^{r_2}K_0\rmd r+\frac{\pi}{2}\right)= &
& \nonumber \\
A\frac{L N_0^\star}{r_\star\omega} 
\sin\left(\int_{r_1}^{r_\star}K_0\rmd  r+\frac{\pi}{4}\right)\sin\left(\int_{r_\star}^{r_2}K_0\rmd
  r+\frac{\pi}{4} \right).
\label{eigengap}
\end{eqnarray}
\normalsize

Equation (\ref{eigengap}) provides us the eigenvalue condition in the
presence of a glitch. 
 We note that this condition differs from those derived following
  similar principles by \cite{brassard92} and \cite{miglio08} for g modes in
  white dwarfs and main-sequence stars, respectively, in particular because 
  we model the glitch by a Dirac delta rather than a step
  function.\footnote{We note that the mathematical derivation of the eigenvalue condition
    in the present work differs substantially from that presented by
    \cite{brassard92} and \cite{miglio08},  in that it is based on a
    single equation for the variable $\Psi$, rather than on the
    equations for variables related to the radial displacement and the
    Eulerian pressure perturbation.
  } 

To write the eigenvalue condition in a form that can be compared with
the one derived without the glitch, we use the relation
\small
\begin{eqnarray}
\int_{r_1}^{r_\star}K_0\rmd r+\frac{\pi}{4}=\int_{r_1}^{r_2}K_0\rmd
r+\frac{\pi}{2}-\int_{r_\star}^{r_2}K_0\rmd r-\frac{\pi}{4}.
\label{int}
\end{eqnarray}
\normalsize
Introducing equation~(\ref{int})  in equation~(\ref{eigengap}) we find, after some algebra,
\begin{eqnarray}
\sin\left(\int_{r_1}^{r_2}K_0\rmd  r+\frac{\pi}{2}+\Phi\right) = 0.
\label{eigengap2}
\end{eqnarray}
In the above, the phase $\Phi$  is defined by the following system of
equations,
\begin{equation}
\left\{
\begin{array}{lll}
B\cos\Phi \hspace{-0.0cm}&\hspace{-0.0cm}= & \hspace{-0.0cm} 1-A\displaystyle{\frac{LN_0^\star}{r_\star\omega}}
  \sin\left(\int_{r_\star}^{r_2}K_0\rmd r+\frac{\pi}{4}
  \right)\times \\
&&\cos\left(\int_{r_\star}^{r_2}K_0\rmd r+\frac{\pi}{4}
  \right) 
\\ \\
B\sin\Phi \hspace{-0.0cm}&\hspace{-0.0cm}= &\hspace{-0.0cm}
A\displaystyle\frac{LN_0^\star}{r_\star\omega}\sin^2\left(\int_{r_\star}^{r_2}K_0\rmd
  r+\frac{\pi}{4} \right),
\end{array}
\right.
\label{Phi_g}
\end{equation}
where $B$ is a function of frequency, also defined by the system of
equations~(\ref{Phi_g}). Thus, we arrive at the final form of the
eigenvalue condition for our toy model when including the glitch in
the buoyancy frequency, namely,
\begin{equation}
\int_{r_1}^{r_2}K_0\rmd r = \pi\left(n-\frac{1}{2}\right)-\Phi.
\label{eigen_glitch}
\end{equation}
By comparing equations~(\ref{eigen}) and (\ref{eigen_glitch}) we see that the 
frequencies of pure g modes are modified by the
glitch through the frequency dependent phase $\Phi$ only.

\subsubsection{Effect on the period spacing}
\label{sec:PS_gmode}
Having considered the effect of the glitch on the g-mode
frequencies, we now turn to the impact it has on the g-mode period
spacing, defined as the difference between the periods of two modes of the same degree
and consecutive radial orders.  A possible way to proceed would be
  to solve the eigenvalue condition  numerically \cite[as done,
  {\it e.g.}, by][]{brassard92, miglio08} to find the oscillation
  frequencies and, thus, compute the period spacings.  Instead, we opt
  for  deriving an analytical expression that directly describes
  the period spacings as a function of the oscillation frequency in
  terms of the glitch parameters, which we find may be a useful path for the future
  comparison with the period spacings derived from real data.

Under the
asymptotic approximation, the period spacing for high-radial
order g  modes,
\dPa, is essentially constant and given by \citep{tassoul80},
\begin{equation}
\Delta P_{\rm as}\simeq \frac{2\pi^2}{\omega_{\rm g}},
\label{psasymp}
\end{equation}
where,
\begin{equation}
\omega_{\rm g}\equiv\int_{r_1}^{r_2}\frac{LN_0}{r}\rmd r.
\label{omegag}
\end{equation}
To see how the period spacing is modified from the asymptotic value in
the presence of the glitch, we
first re-write the eigenvalue condition~(\ref{eigen_glitch}) as, 
\begin{equation}
\frac{\omega_{\rm g}}{2\pi}P+\Phi\approx \pi\left(n-\frac{1}{2}\right),
\label{Gfunction}
\end{equation}
where $P=2\pi/\omega$ is the oscillation period (and we recall that
$\Phi$ is itself a function of $P$). In deriving the above, we have used the fact that well within the g-mode cavity
$K_0\approx LN_0/\omega r$ to approximate $\int_{r_1}^{r_2}K_0\rmd r$ by
$\int_{r_1}^{r_2}{LN_0}/{\omega r\,}\rmd r$. Because $K_0$ goes to
zero towards the turning points, this
approximation leads to a slight overestimate of the value of the
integral. However, it allows us to derive a simple analytical
expression for the period spacing.

Next, we follow \cite{jcd12}\footnote{Note that the analysis of the simplified model discussed in Section 4.2 of that paper contains 
two errors that fortuitously cancel. One is the neglect of a singularity in the asymptotic expression 
(equation (1) of that paper) in the evanescent region. The second is a simple sign error in the analysis leading to equation (22) of that paper. 
The combined effect of the errors is that the equation has the correct form, 
and the remaining analysis is still valid. } and define a function $\mG (P)$, by
\begin{equation}
\mG\left(P\right) = \frac{\omega_{\rm g}}{2\pi}P+\Phi.
\end{equation}
 Using expression~(\ref{Gfunction}) and the definition of $\mG$, we can
then write
\begin{equation}
\pi\approx \mG\left(P_{n+1}\right)-\mG\left(P_{n}\right)\approx\frac{\rmd\mG}{\rmd P}\Delta P ,
\label{dgdp}
\end{equation}
where $\Delta P = P_{n+1}-P_{n}$ is the period spacing in the presence of the glitch, 
or, equivalently,
\begin{equation}
\Delta P\approx\frac{\pi}{\displaystyle\frac{\rmd\mG}{\rmd P}}.
\label{dgdp_2}
\end{equation}
By differentiating $\mG$ with
respect to $P$ and substituting in equation~(\ref{dgdp_2}) we find that this period spacing is related to the asymptotic period spacing by
\begin{equation}
\Delta P\approx \frac{\Delta P_{\rm as}}{1-\displaystyle\frac{\omega^2}{\omega_{\rm
    g}}\frac{\rmd\Phi}{\rmd\omega}} \equiv 
    \frac{\Delta P_{\rm as}}{1-\mF_{\rm G}}.
\label{ps_glitch}
\end{equation}
The deviation of the period spacing from its asymptotic
value is reflected in the term $\mF_{\rm G}$.  Its dependence on the glitch
parameters can be made explicit by solving the system of
equations~(\ref{Phi_g}) .  Defining,
\begin{eqnarray}\omega_{\rm g}^\star \equiv
\int_{r_\star}^{r_2}\frac{LN_0}{r}\rmd r 
\label{omegag_star}
\end{eqnarray}
and making $\int_{r^\star}^{r_2}K_0\rmd r \approx \omega_{\rm g}^\star $ we find,
\begin{eqnarray}
\hspace{-1.cm}\mF_{\rm G}\hspace{-0.0cm} & = \hspace{-0.0cm}&
\frac{ALN_0^\star}{r_\star\omega_{\rm
      g}B^2}\left\{\frac{\omega_{\rm g}^\star}{\omega} \cos\left(2 \frac{\omega_{\rm
          g}^\star}{\omega}\hspace{-0.00cm}\right)\right. \nonumber\\
& &\hspace{1cm}+\left.\left(1-\frac{ALN_0^\star\omega_{\rm g}^\star}{r_\star\omega^2}\right)\sin^2\left(
      \frac{\omega_{\rm
          g}^\star}{\omega}\hspace{-0.00cm}+\hspace{-0.00cm}\frac{\pi}{4}\hspace{-0.00cm}\right)\right\},
\label{fg}
\end{eqnarray}
\normalsize
and 
\begin{eqnarray}
\hspace{-2.2cm}  B^2 &= &\left[1-\frac{ALN_0^\star}{2 r_\star\omega}\cos\left(2 \frac{\omega_{\rm
          g}^\star}{\omega}\hspace{-0.05cm}\right)\right]^2\nonumber\\
&&\hspace{2.2cm} +\left[\frac{ALN_0^\star}{r_\star\omega}\sin^2\left(
      \frac{\omega_{\rm
          g}^\star}{\omega}\hspace{-0.05cm}+\hspace{-0.05cm}\frac{\pi}{4}\hspace{-0.05cm}\right)\right]^2.
\label{b2}
\end{eqnarray}
\normalsize
In the above, the dependance of the period spacing on the characteristics of the
glitch is expressed by the parameters $A$ (glitch strength) and
$r_\star$ (glitch position).

   \begin{figure}
\centering
 \includegraphics[width=8.5cm]{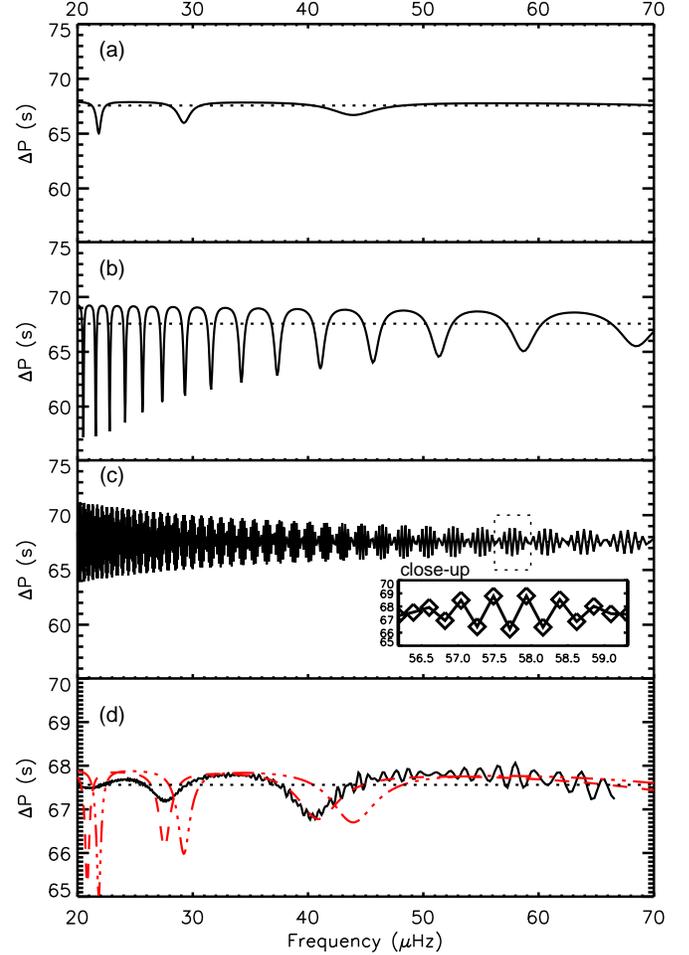}
\vspace{-0.8cm}
     \caption{Period spacing for pure g modes in model~1a. The horizontal dotted line shows the glitch-free period spacing,
       $\Delta P_{\rm as}$.  {\bf (a)} Results derived from
       expression~(\ref{ps_glitch})  for the glitch parameters
       estimated for model~1a, $r_\star=0.02164R$ (or $\omega_{\rm
         g}^\star/\omega_{\rm g}=0.0057)$) and $A=1.8\times
       10^{-3}R$ (see text for details). {\bf (b)} Results from
       expression~(\ref{ps_glitch}),
       but for a slightly deeper glitch at $\omega_{\rm
         g}^\star/\omega_{\rm g}=0.0275$.  {\bf (c)} Results from
       expression~(\ref{ps_glitch}),
       for an even deeper glitch, at $\omega_{\rm
         g}^\star/\omega_{\rm g}=0.4600$. The inset is a close-up
       of the region enclosed by the dashed box. Diamonds show the individual 
      modes. See text for details. 
      {\bf(d)} Results obtained from integrating
       the wave equation numerically, ignoring the coupling to the
       p modes (solid black curve). Overplotted is the result derived from
       expression~(\ref{ps_glitch})  for the glitch parameters
       estimated for model~1a  (dotted-dashed red curve; same as solid
       black line in panel a) and the result obtained from the same
       expression with the glitch parameters 
       adjusted to the numerical solution, namely $r_\star=0.0221R$  (or $\omega_{\rm
         g}^\star/\omega_{\rm g}=0.0052$) and $A=1.65\times 10^{-3}R$ (red, dashed
       curve). The latter has been shifted in frequency by
       1$\muHz$ (see text
       for details).
  }
     \label{PS_gmode}
   \end{figure}

The period spacing derived from
expression~(\ref{ps_glitch}) for our model~1a is
illustrated in Figure~\ref{PS_gmode}a (solid curve).  It varies around the
asymptotic value (horizontal dotted line), forming
relatively narrow dips that alternate with wider, less pronounced
humps. The narrowing of the dips with decreasing frequency is due to
the $1/\omega$ dependence of the arguments of the sinusoidal functions
present in  expressions~(\ref{fg}) and (\ref{b2}).\footnote{
We emphasize that unlike the case
of the dips caused by mode coupling, the glitch-induced dips are not
associated with the presence of an extra mode. Thus, the decrease in
the period spacing at the dips is fully compensated by its
increase at the wider, less-pronounced humps.}

Because all sinusoidal functions
present in the definition of  $\mF_{\rm G}$ and $B$
(equations~(\ref{fg}) and (\ref{b2}), respectively)
can be written in terms of the argument $2\omega_{\rm g}^\star
/\omega$ ($\equiv \pi^{-1}\omega_{\rm g}^\star P$), we expect the distance between 
dips to be constant in period and equal to
$2\pi^2/\omega_{\rm g}^\star$. Thus, it  provides a measure of the depth of
the glitch in terms of the normalized buoyancy depth,
\begin{equation}
\frac{\omega_{\rm g}^r}{\omega_{\rm g}}\equiv\frac{1}{\omega_{\rm
    g}}\int_{r}^{r_2}\frac{LN_0}{r}\rmd r,
\label{bdepth}
\end{equation}
which is analogous to the normalized acoustic depth used in
studies of acoustic waves\footnote{Here we adopted the notation of
\cite{montgomery03}, where the buoyancy depth is defined as the inverse
  of a period, resulting in the sinusoidal part of the eigenfunction
  having approximately the form $\sin(\pi^{-1}\omega_{\rm g}^r P+\pi/4)$.  However, we note that the term buoyancy depth
  is sometimes used for $L/\displaystyle{\omega_{\rm  g}^r}$ \citep[e.g.][]{miglio08},
instead.}. Down to the middle of the cavity (located at $\omega_{\rm  g}^r/\omega_{\rm g}=0.5$), the
deeper the glitch location, the smaller the spacing between dips. For yet deeper glitches
($\omega_{\rm  g}^r/\omega_{\rm g} > 0.5$), the spacing between dips
increases again, mirroring the separation found for a glitch
positioned at $1-\omega_{\rm  g}^r/\omega_{\rm g}$ \citep[e.g.][]{montgomery03}.

For model~1a, if we
take $r_\star=0.02164R$ (the radius at which  $N^2-N_0^2$  is maximum) we find
$\omega_{\rm g}^\star/\omega_{\rm g}=0.0057$. Hence the glitch is very
close to the the edge of the cavity when measured in terms of
$\omega_{\rm  g}^r/\omega_{\rm g}$ (equation~(\ref{bdepth})).

Figures~\ref{PS_gmode}b and \ref{PS_gmode}c show the results of moving the glich
deeper inside the cavity, to $\omega_{\rm g}^\star/\omega_{\rm g}=0.0275$ and $\omega_{\rm
  g}^\star/\omega_{\rm g}=0.4600$, respectively. As expected, the
spacing between the dips at fixed frequency
gets smaller as the glitch is moved closer to the center of the
cavity \cite[see also figures 8 and 15 of][ which show a similar
effect for the g-mode period spacings in main-sequence classical pulsators]{miglio08}.  We note that in producing
Figures~\ref{PS_gmode}b and \ref{PS_gmode}c we have also changed
$A$ from the value used in Figure~\ref{PS_gmode}a. In
Figure~\ref{PS_gmode}b, $A$ was chosen such as to maintain the value
of the {\it effective} glitch strength $\tilde
A\equiv ALN_0^\star/r_\star$ (see right hand side of
condition~(\ref{eigengap}))  unchanged. Thus, the difference in the
amplitudes of the patterns seen in Figures ~\ref{PS_gmode}a and
\ref{PS_gmode}b results solely from the difference in the location of
the glitch. For Figure~\ref{PS_gmode}c, $A$ was
chosen such as to reduce the effective strength by one
order of magnitude. In this limit of small effective strength the
period spacing shows symmetric wiggles around the
asymptotic value, instead of the alternating
dips and humps seen in the other two cases.  Interestingly, in  Figure~\ref{PS_gmode}c we can
identify a modulation of the period spacing on a scale larger than the
separation between wiggles. This modulation is more noticeable when the distance
between adjacent modes becomes comparable
with the distance between glitch-induced minima. It is simply a
sampling effect, as can be confirmed through inspection of the
inset of Figure~\ref{PS_gmode}c. We note, however, that for a glitch positioned at
$\omega_{\rm g}^\star/\omega_{\rm g}=0.5$, the period spacing between two minima is
exactly twice the asymptotic period spacing, creating a perfect
sawtooth diagram without the modulation seen in
Figure~\ref{PS_gmode}c. 
The modulation introduced by the limited sampling depends solely on $\omega_{\rm
  g}^\star/\omega_{\rm g}$, thus providing an alternative way to
measure the position of the glitch. This is important,
because due to the limited frequency resolution of the observations, it
might, in some stars, be easier to detect this larger scale
modulation than the series of glitch-induced
variations in the period spacing

Since in reality the glitch is not infinitely narrow, estimating the parameters
$r_\star$ and $A$ from a given model requires a little
thought. To estimate $r_\star$ one may consider taking either the center of the glitch or the position of its
maximum amplitude. However, to estimate $A$ we need to
consider how to transform the glitch in the stellar model into its
infinitely narrow counterpart while
keeping the area under the glitch essentially unchanged.  Recalling that the Dirac $\delta$ can be defined as the limit,
\begin{equation}
\delta = \lim_{\epsilon \to 0^+}\frac{1}{\epsilon\sqrt\pi}{\rm e}^{-\left(r-r_\star\right)^2/\epsilon^2},
\end{equation}
and taking $\epsilon$ to be the characteristic half width of the
glitch we find, from equation~(\ref{glitch}),
\begin{eqnarray}
\left.\frac{\Delta{N^2}}{N_0^2}\right|_{r_\star}\approx
\frac{A}{\epsilon\sqrt\pi},
\label{amplitude}
\end{eqnarray}
where $\Delta N^2 = N^2-N_0^2$ is the glitch induced deviation in the
square of the buoyancy frequency. Taking $r_\star$ to be the  radius 
at which  $\Delta N^2$  is maximum,  we estimate that
$\epsilon=0.5 \times 10^{-3}R$ and $A=1.8\times 10^{-3}R$,  for our model~1a.

\subsubsection{Numerical solution for pure g modes}
\label{gnumerical}
In the next step we will move to a more realistic description of the
effect from a glitch on the period spacing. Figure~\ref{figeigen}  shows
that a Dirac $\delta$ function is not a realistic description of the
glitch in our stellar model.  In principle, the analytical
analysis could  include a more realistic function to describe the
glitch. However, that would have increased the complexity of the
analysis  whose
main purpose was to provide a simple understanding of the seismic impact of
the glitch. To obtain more realistic results we therefore solve
equation~(\ref{waveeqap}) numerically, for the
case of pure g modes, by adopting $N$ from the stellar
structure model.  By comparing the results  with those derived
analytically, we can investigate the impact of the approximations made in the analytical
analysis and produce results that are more
directly comparable  with the full numerical solutions from \adipls\
that will be discussed in section~\ref{fullnumerical}.

To find the numerical solutions for pure g modes we approximate
$K^2$ in equation~(\ref{waveeqap}) by 
\begin{eqnarray}
K^2=-\frac{L^2}{r^2}\left(1-\frac{N^2}{\omega^2}\right).
\label{k2_gmode}
\end{eqnarray}
The equation is then solved using a standard fourth order Runge-Kutta method with adaptive step size control and the eigenfrequencies are found by
imposing that the solutions satisfy the boundary conditions $\Psi=0$
at $r=0$ and $r=R$. 

The results are
presented in Figure~\ref{PS_gmode}d (solid, black curve).  Comparison with the analytical
results derived in section~\ref{sec:PS_gmode} for the glitch parameters
estimated for model~1a (Figure~\ref{PS_gmode}a; also shown as
dotted-dashed, red curve in Figure~\ref{PS_gmode}d) provides a number of interesting conclusions. 
First, and most importantly, the general form of the period spacing variation is
similar in the two cases, reemphasizing that the
effect of the glitch is the formation of narrow dips that alternate with
wider, much less pronounced humps. However, it is also clear that both the depth of the dips and their
  separation in frequency are different in the analytical and numerical
results. To understand these differences and their potential impact on glitch-parameter
inferences based on the analytical model we adjust the
glitch parameters such as to match the analytical to the numerical
results.  The new analytical solution is shown by the red-dashed
curve in Figure 4d. The solution had to be shifted in frequency by 1$\mu$Hz
because the approximation $\int_{r^\star}^{r_2}K_0\rmd r \approx
\omega_{\rm g}^\star $ made earlier introduces a phase shift between
the analytical and the numerical results.  In practice, this may be accounted for
by adding a phase to the arguments of the sinusoidal functions in the
analytical model, thus,  increasing the number of adjustable parameters by one.

The rematched
glitch location, $r_\star$, is almost unchanged (shifted by only $\approx 50\%$
of the glitch width), while the strength of the glitch is about $10\%$
smaller. 
The latter reflects that  the period spacing variations
have a lower amplitude in the numerical results.
This difference in the amplitudes and, more
notably, the fact that they vary in opposite ways with frequency, is a consequence of the
non-negligible width of the glitch. Towards lower frequency 
the g-mode wavelength becomes shorter.  Seen by the wave, a spike in
the \brunt therefore appears smoother (less of a glitch). As a result,
the amplitude of the dips in the period spacing becomes
smaller towards lower frequency. However, in the analytical
analysis the spike is modelled as being infinitely narrow. It is therefore always much narrower than the local wavelength
and, hence, no reduction of the amplitude is seen.  

A second striking difference seen in Figure~\ref{PS_gmode}d concerns
the small-scale variations that are present in the numerical
result, but absent in the analytical curve. Using our analytical
model (equation~(\ref{ps_glitch})) we found that these small-scale variations
would originate from a glitch at the hydrogen-burning shell.  Given that the spike in
the \brunt at this position is not seen as a glitch by the wave (as discussed in section~\ref{structure}) we inspected the derivatives
of $N$ and found that they show a high level of variation at much
smaller scales than the local wavelength.   By smoothing the derivatives
and recalculating the period spacing, the small-scale variations disappeared. We therefore conclude that
their origin is purely numerical and has no
physical meaning.

\subsection{Coupling with the p modes}
\label{coupling}

We now consider the same problem as in section~\ref{ganalytical}, but
include the coupling between the
g and p modes.  That requires replacing
solution~(\ref{gasymp_r}), valid for pure g modes, by the solution that accounts for mode
coupling.  

When we  consider that waves can propagate also in the
p-mode cavity,  the asymptotic solution to equation~(\ref{waveeqap})
that is valid well within the evanescent region is no longer an exponentially decaying function,
but rather a linear combination of an exponentially decaying and an exponentially
growing function.  The solution to equation~(\ref{waveeqap})
that matches the required linear combination has the form \citep{gough93},
\begin{equation}
\Psi_{\rm out} \sim \hat{\Psi}_{\rm out} K_0^{-1/2} \sin\left(\int_r^{r_2} K_0\rmd r +
  \frac{\pi}{4}+ \varphi\right), 
\label{gasymp_r_c}
\end{equation}
 in the region $ r_1 \ll r \ll r_2$,
where $\hat{\Psi}_{\rm out}$ is a constant and $\varphi$ is a frequency dependent phase, which is uniquely
defined by the coefficients of the linear combination mentioned above. Its form will be discussed 
below.

Equations~(\ref{gasymp_l}) and (\ref{gasymp_r_c}) provide us the
eigenvalue condition in the presence of mode coupling and no glitch, namely,
\begin{equation}
\int_{r_1}^{r_2}K_0\rmd r = \pi\left(n-\frac{1}{2}\right)-\varphi.
\label{eigen_coupling}
\end{equation}
The coupling phase $\varphi$ can be obtained from the
eigenvalue condition derived by \cite{shibahashi79} (see
also~\cite{unno89}), based on an asymptotic analysis of  the equations
for the radial component of the displacement and for the Eulerian pressure perturbation, under
the Cowling approximation. 
Because the oscillation frequencies must be independent of the variable used to express
the pulsation problem, the eigenvalue condition derived by \cite{shibahashi79} must be
equivalent to our eigenvalue
condition~(\ref{eigen_coupling}). Comparing the two we find (see
appendix~A, for details),
\begin{equation}
\varphi \approx {\rm atan}\left[\frac{q}{\tan\left(\displaystyle\frac{\omega -\omega_{\rm a}}{\omega_{\rm p}}\right)}\right],
\label{varphi}
\end{equation}
where $q$ is a frequency dependent coupling factor that can take
values in the  range $ 0 \le q < 1/4$, where
smaller values imply a weaker coupling. Moreover, $\omega_{\rm a}$ are the oscillation frequencies that would be obtained for
p modes in the absence of coupling (the acoustic resonant frequencies) and $\omega_{\rm
  p}=\left(\int_{r_3}^{r_4}c^{-1}\rmd r\right)^{-1}$ is approximately
twice the asymptotic large separation. The corresponding period
spacing, derived as in section~\ref{sec:PS_gmode}, is given by,
\begin{equation}
\Delta P\approx \frac{\Delta P_{\rm as}}{1-\displaystyle\frac{\omega^2}{\omega_{\rm
    g}}\frac{\rmd\varphi}{\rmd\omega}}\equiv 
    \frac{\Delta P_{\rm as}}{1-\mF_{\rm C}}.
\label{ps_coupling}
\end{equation}
    \begin{figure*} 
          \begin{minipage}{0.5\linewidth}
          \hspace{-0.06\linewidth}
               \rotatebox{270}{\includegraphics[width=0.9\linewidth]{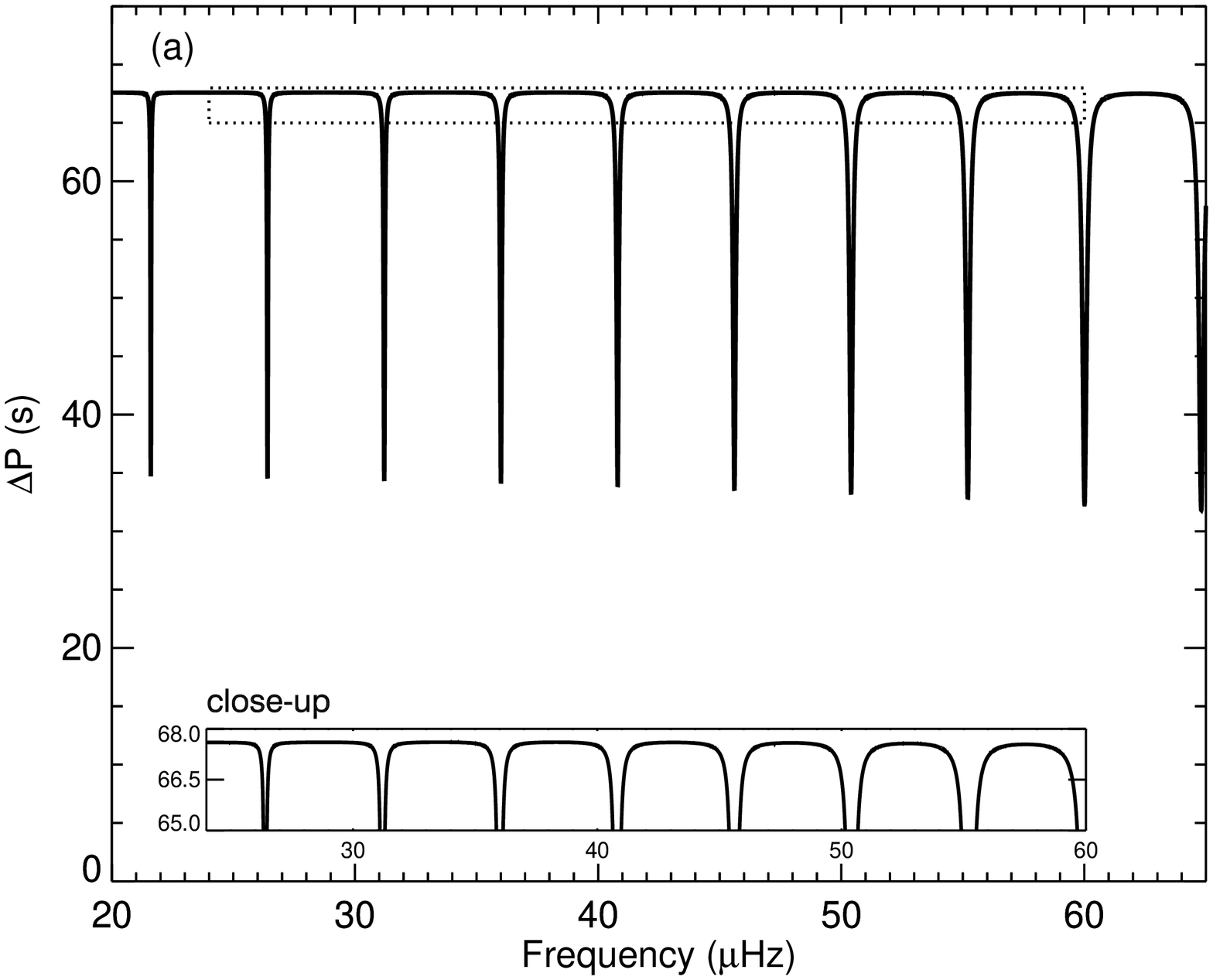}}
          \end{minipage}
          \begin{minipage}{0.5\linewidth}
            \hspace{-0.04\linewidth}
             \rotatebox{270}{\includegraphics[width=0.9\linewidth]{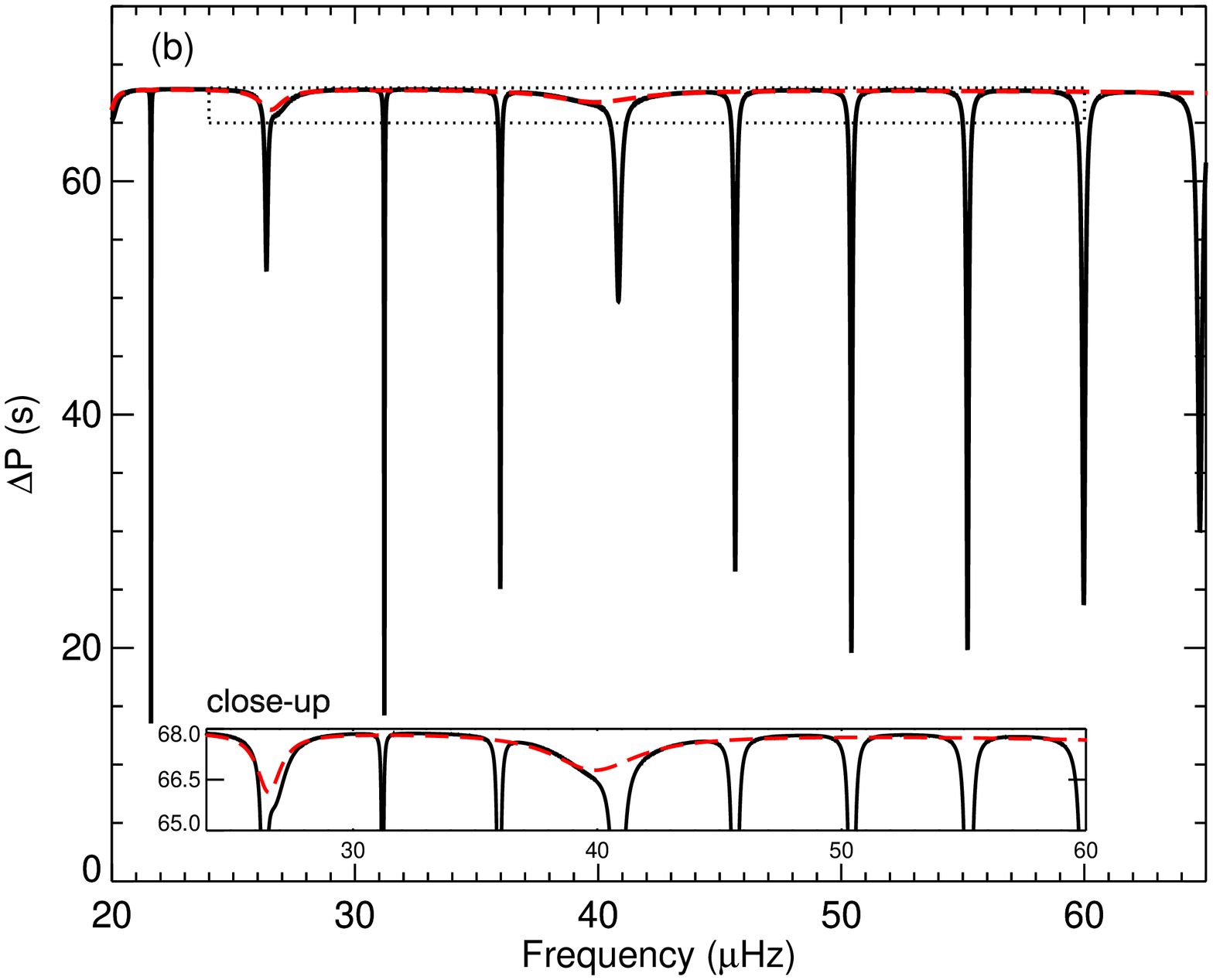}}
           \end{minipage}
        \caption{Period spacing derived from the analytical approach
          for model~1a. The inset shows a close-up
          around $\Delta P_{\rm as}$ (dotted box). {\bf (a)} case with mode
        coupling but no glitch, computed from
        expression~(\ref{ps_coupling}). {\bf (b)} case with
        mode coupling and glitch, computed from
        expression~(\ref{ps_coupling_glitch}) (black) and case
        with glitch and no mode coupling, computed from
        expression~(\ref{ps_glitch}) (red), adopting $r_\star=0.0221R$  and $A=1.65\times 10^{-3}R$. }
         \label{ps_an_coup}

          \end{figure*}

The period spacing derived from expression (\ref{ps_coupling}) for our
model~1a is shown in
Figure~\ref{ps_an_coup}a. The dips associated with the coupling to the
p modes are
equally spaced in frequency and located at the acoustic resonant (cyclic) 
frequencies ($\omega_{\rm a}/2\pi$). At these frequencies the denominator inside
the arctan of~(\ref{varphi}) goes through zero and, as a consequence,
$\varphi$ varies rapidly with frequency. The large
derivative in frequency of $\varphi$ therefore makes $\mF_{\rm C}$ large, producing the
dips in the period spacing. This is in agreement with the
discussions by \cite{jcd12} and \cite{mosser12b} and with the period spacing derived from
the analysis of real data
for red-giant stars \citep[e.g.][]{beck11}.

Next, we add the effect of the glitch. Following the same
steps as in section~\ref{ganalytical}  we find that the eigenvalue condition
in the presence of mode coupling and a glitch is given by
\begin{equation}
\int_{r_1}^{r_2}K_0\rmd r = \pi\left(n-\frac{1}{2}\right)-\Phi-\varphi,
\label{eigen_glitch_coupling}
\end{equation}
where $\Phi$ is now defined by the following system of equations,
\begin{equation}
\left\{
\begin{array}{lll}
B\cos\Phi \hspace{-0.0cm}&\hspace{-0.0cm}= & \hspace{-0.0cm} 1-A\displaystyle\frac{LN_0^\star}{r_\star\omega}
  \sin\left(\int_{r_\star}^{r_2}K_0\rmd r+\frac{\pi}{4}+\varphi
  \right) \times\\ 
& & \cos\left(\int_{r_\star}^{r_2}K_0\rmd r+\frac{\pi}{4}+\varphi
  \right) 
\\ \\
B\sin\Phi \hspace{-0.0cm}&\hspace{-0.0cm}= &\hspace{-0.0cm} A\displaystyle\frac{LN_0^\star}{r_\star\omega}\sin^2\left(\int_{r_\star}^{r_2}K_0\rmd
  r+\frac{\pi}{4} +\varphi\right).
\end{array}
\right.
\label{Phi_coupling}
\end{equation}
We emphasize that both $B$ and $\Phi$ now depend on $\varphi$. This is to be expected,
since the effect of the glitch on the oscillations depends critically on
the phase of the eigenfunction at the depth where the glitch is
located, and that phase is influenced by the coupling.  As a
consequence, the relative deviation of the period spacings from
  the asymptotic value when both a glitch and mode coupling are present  is different from  what
would be found by simply adding the deviations
generated by the coupling and by the glitch separately.  This fact can be readily
seen in the period spacing derived from the eigenvalue
condition~(\ref{eigen_glitch_coupling}), which has the form
\begin{equation}
\Delta P\approx \frac{\Delta P_{\rm as}}{1-\displaystyle\frac{\omega^2}{\omega_{\rm
    g}}\left[\frac{\rmd\Phi}{\rmd\omega}+\frac{\rmd\varphi}{\rmd\omega}\right]}\equiv 
    \frac{\Delta P_{\rm as}}{1-\mF_{\rm {G,C}}}.
\label{ps_coupling_glitch}
\end{equation}
As before, the deviation of the period spacing from its asymptotic
value is reflected in the second term, $\mF_{\rm {G,C}}$, present in the denominator of
expression~(\ref{ps_coupling_glitch}).  Its dependence on the glitch
parameters can be made explicit by solving the system of
equations~(\ref{Phi_coupling}), from which we obtain
\begin{eqnarray}
\hspace{-1.cm}\mF_{\rm {G,C}}\hspace{-0.0cm} &=& \nonumber\\
& &  \hspace{-1.cm}\frac{\omega^2}{\omega_{\rm
    g}}\frac{\rmd\varphi}{\rmd\omega}\left\{1+\frac{ALN_0^\star}{r_\star\omega
    B^2}\left[\cos\left(2 \frac{\omega_{\rm
          g}^\star}{\omega}\hspace{-0.00cm}+\hspace{-0.00cm}
      2\varphi\right)\right.\right. \nonumber \\ 
&& \hspace{1.5cm}\left.\left. -\frac{ALN_0^\star}{r_\star\omega}\sin^2\left(
      \frac{\omega_{\rm
          g}^\star}{\omega}\hspace{-0.00cm}+\hspace{-0.00cm}\frac{\pi}{4}\hspace{-0.00cm}+\hspace{-0.00cm}\varphi\right)\right]\right\}
\nonumber \\ 
 &&\hspace{-1.cm} +\hspace{-0.0cm} \frac{ALN_0^\star}{r_\star\omega_{\rm
      g}B^2}\left\{\frac{\omega_{\rm g}^\star}{\omega} \cos\left(2 \frac{\omega_{\rm
          g}^\star}{\omega}\hspace{-0.00cm}+\hspace{-0.00cm}
      2\varphi\right)\right. \nonumber \\
&&\hspace{0.1cm}\left.+\left(1-\frac{ALN_0^\star\omega_{\rm g}^\star}{r_\star\omega^2}\right)\sin^2\left(
      \frac{\omega_{\rm
          g}^\star}{\omega}\hspace{-0.00cm}+\hspace{-0.00cm}\frac{\pi}{4}\hspace{-0.00cm}+\hspace{-0.00cm}\varphi\right)\right\},
\label{fgc}
\end{eqnarray}
\normalsize
where, $B^2$ is now given by
\begin{eqnarray}
\hspace{-2.0cm} B^2 \hspace{-0.05cm}&= &\hspace{-0.05cm}\left[1 \hspace{-0.05cm}-\hspace{-0.05cm}\frac{ALN_0^\star}{2 r_\star\omega}\cos\left(2 \frac{\omega_{\rm
          g}^\star}{\omega}\hspace{-0.05cm}+\hspace{-0.05cm}
      2\varphi\right)\right]^2 \nonumber \\
&&\hspace{2.cm}+\hspace{-0.05cm}\left[\frac{ALN_0^\star}{r_\star\omega}\sin^2\left(
      \frac{\omega_{\rm
          g}^\star}{\omega}\hspace{-0.05cm}+\hspace{-0.05cm}\frac{\pi}{4}\hspace{-0.05cm}+\hspace{-0.05cm}\varphi\right)\right]^2.
\label{b2gc}
\end{eqnarray}
\normalsize

We see that $\mF_{\rm {G,C}}$ has two terms, each marked by a set of curly brackets.
If we assume there is no coupling, meaning that $\varphi$  is zero
for all frequencies, the first term vanishes because
$\rmd\varphi/\rmd\omega = 0$, and the second term becomes equal to
equation~(\ref{fg}), hence reducing  $\mF_{\rm {G,C}}$ to $\mF_{\rm G}$ as
expected.  If instead we assume there is no glitch, which
translates to $A=0$, then the second term vanishes, and the first term
becomes equal to $\mF_{\rm {C}}$ defined in equation~(\ref{ps_coupling}), reducing $\mF_{\rm {G,C}}$ to  $\mF_{\rm {C}}$; again as one would expect.
Finally, we consider that both coupling and glitch are present,
but we look specifically at what happens at the acoustic
resonance where the coupling dominates the expression for $\mF_{\rm {G,C}}$.
Here, the frequency derivative of $\varphi$  is very large, and hence the
first term is generally much larger than the second. However, we still have the two extra $\cos$ and $\sin^2$ `glitch-induced' terms within the first set of curly brackets compared to just $\mF_{\rm {C}}$ ($=\omega^2/\omega_{\rm
    g} \,\, \rmd\varphi/\rmd\omega$). This shows that the mode coupling, and therefore also the dips located at the acoustic resonance frequencies, is influenced by the glitch.

The  period spacing obtained from
expression~(\ref{ps_coupling_glitch}) is
 illustrated in Figure~\ref{ps_an_coup}b. Comparing with Figure~\ref{ps_an_coup}a, we see
that the combined effect of the glitch and the coupling on the
period spacing is predominantly a change in the depth of the
dips at the acoustic resonant frequencies. Whenever a dip 
caused by the glitch coincides with a dip caused by the coupling with
the p modes, the depth of the latter is reduced. But if a hump
produced by the glitch coincides with the dip caused by
the coupling,  the depth of the dip increases.  This
behaviour is oposite to what would be found if the the combined effect
were simply the sum of the deviations to the asymtotic period
spacings caused by each effect separatly.
The predicted behaviour  can be understood if we recall that the extent to which g
modes couple to a p mode depends critically on the proximity of their
frequencies (assuming everything else remains unchanged, which is the
case here).
A glitch-induced dip in the period spacings means the g modes are
locally more
densely packed, as compared to the asymptotic
case. Thus, if the dip coincides with an acoustic resonant frequency
the number of g modes coupling to the p mode is greater, resulting
in a wider and, consequently, shallower coupling dip. However, if an acoustic resonant frequency
coincides with a glitch-induced hump, the number of g modes coupling
to the p mode is reduced, resulting in a thinner, hence, deeper
coupling dip.

\section{Interpretation of full numerical solutions}
\label{fullnumerical}

In this section we consider the numerical solution of the full
pulsation equations, including the perturbation to the gravitational
potential, for the models introduced in section~\ref{structure} and
interpret them in the light of the results found with the toy model analysis
presented in section~\ref{g-modes}. The full numerical
solutions were computed with the pulsation code \adipls. Care was taken to have an adequate number 
of mesh points with appropriate distribution to resolve the rapidly varying eigenfunctions in 
the g-mode cavity.

\begin{figure} 
          \hspace{-0.06\linewidth}
              \rotatebox{270}{\includegraphics[width=0.9\linewidth]{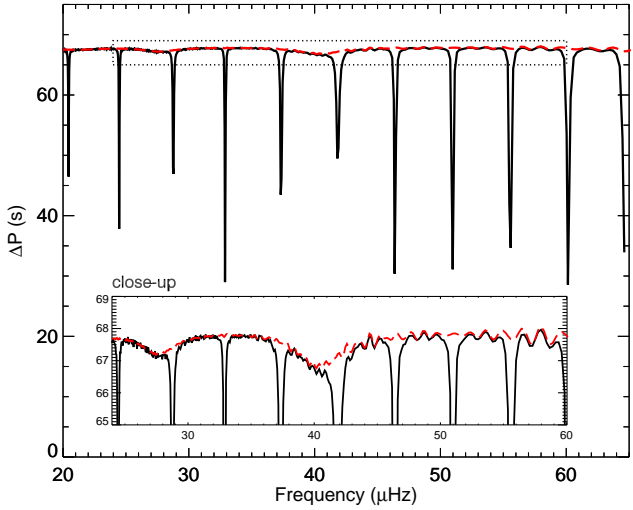}}
          \caption{ Period spacings derived from
            \adipls\ for model 1a, including the effects of the glitch and of the
            coupling between the g-modes and the p-modes (solid curve), compared to the integration of the wave
            equation (\ref{waveeqap}), ignoring the coupling with
            p modes (red, dashed curve).  The inset shows a close up
          around $\Delta P_{\rm as}$ (dotted box).}
         \label{ps_full}

          \end{figure}

The period spacing derived for our model~1a from the full numerical
solutions is shown in Figure~\ref{ps_full}  (solid curve). Comparing
with Figure~\ref{ps_an_coup} we see that the dips associated with the
acoustic resonant frequencies are closer in frequency in the results from the full
numerical solutions than in the toy model. This reflects the fact that the large separation
computed from the eigenfrequencies is smaller (by about 5$\%$ in the
current case) than the corresponding asymptotic value, in accordance
with the results of previous studies \cite[e.g.][]{stello09,belkacem13,mosser13}.  Letting aside that
difference, we see that in the full numerical solutions the dips associated with the acoustic resonant frequencies show a depth variation resembling what is
 seen using our toy model (Figure~\ref{ps_an_coup}b). 
Comparison with the period
spacing derived numerically considering only the g modes
(section~\ref{gnumerical}) (red, dashed curve in
Figure~\ref{ps_full}) confirms that the glitch in the buoyancy frequency is the cause of the larger-scale modulation seen in the
full solution (see inset), and, thus, that the combined effect of the
glitch and mode coupling is to reduce/increase the depth of the dips
positioned at the acoustic resonant frequencies when they coincide
with a glitch-induced dip/hump. 

\begin{figure}
\includegraphics[width=8.5cm]{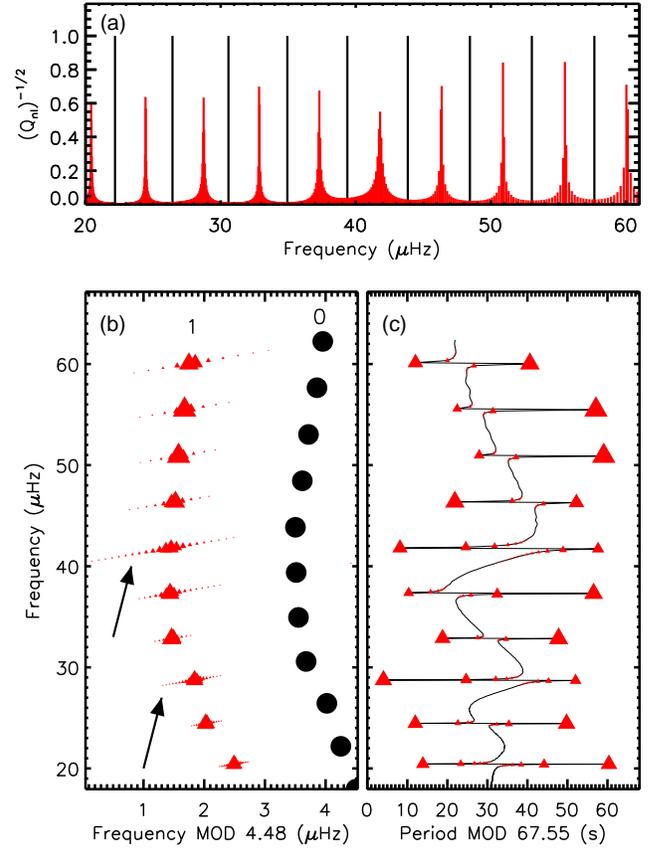}
\caption{Model 1a (\numax\ $ = 45\,$\muhz) -  
(a) Pseudo amplitude spectrum based purely on mode inertia relative to the radial modes
($1/\sqrt{Q_{nl}}$), (radial in black and
dipole in red). 
(b) \'Echelle diagram including radial (circles) and dipole 
(triangles) modes. The abscissa is the frequency modulo \dnu. Symbol size follows the peak heights from panel (a).  The
arrows indicate clusters of dipole modes affected by a glitch-induced dip
in the period spacing. 
(c) Period \'echelle following the notation of panel (b). The abscissa is the period modulo \dPa. The solid curve
connects all the dipole modes. 
\label{echelle}} 
\end{figure}

To illustrate how a glitch in the buoyancy frequency could be revealed in
observational data, Figure~\ref{echelle} shows (a) $1/\sqrt{Q_{nl}}$
representative of relative mode amplitude \citep{jcd04,aerts10}, and
both (b) a frequency- and (c) a period-\'echelle diagram corresponding to
model~1a. Here, $Q_{nl}$ is a measure of the inertia of a mode of
radial order $n$ and degree $l$, relative
to that of radial modes defined by
\begin{equation}
Q_{nl}=\frac{{I}_{nl}}{\overline{{I}}_{l=0}},
\end{equation}
where $I_{nl}$ is the surface normalized mode inertia,
\begin{equation}
I_{nl}=\frac{\int_0^{R_{\rm s}}\left[\xi_{\rm r}^2+l\left(l+1\right)\xi_{\rm
    h}^2\right]\rho r^2\rmd r}{M\xi_{\rm r}\left(R_{\rm s}\right)^2},
\end{equation}
and $\overline{{I}}_{l=0}$ is obtained by interpolating
${I}_{n0}$ to the frequency of the mode under consideration.
Moreover, $\xi_{\rm r}$ and $\xi_{\rm h}$ are the radial and horizontal
components of the displacement, respectively, $M$ is the stellar mass, and
$R_{\rm s}$ is the surface radius.
The frequency \'echelle shows the frequency spectrum
(Figure~\ref{echelle}a) divided into segments of fixed length that are
stacked one above the other. The length of segments equals the average
frequency separation between overtone radial modes, \dnu, found as the
slope of a linear fit to the radial modes versus their order
\citep{grec83}. In the period-\'echelle diagram we show only the dipole modes, and here
the abscissa is the mode period modulo the asymptotic period spacing,
\dPa (equation~(\ref{psasymp}), see e.g., \cite{bedding11}).
 For clarity, we show only modes of relative amplitude above
5\% of the radial modes in the \'echelle diagrams.
Noise set aside, the result seen in this model is a broadening of the clusters of
`observable' dipole modes where the location of the dips coincides with that
of a cluster.  We see this effect in Figure~\ref{echelle}a near 28\muhz\ and
41\muhz\ (see also Figure~\ref{ps_full}).  
In the period-\'echelle diagram, the same effect shows as a strong distortion of
the usual `S' shaped mode pattern seen between each radial mode order in a
glitch-free case (see for example figure~1 of \cite{bedding11}).

\begin{figure} 
            \hspace{-0.06\linewidth}
            \rotatebox{270}{\includegraphics[width=0.9\linewidth]{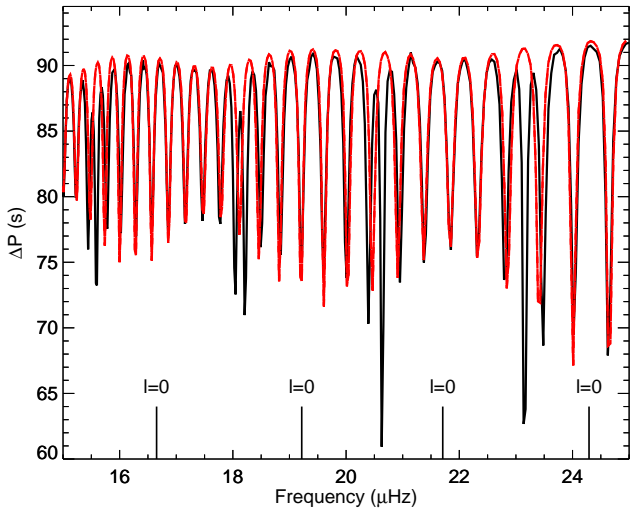}}
          \caption{
            {Period spacings derived from
            \adipls\ for model 1b, including the effects of the glitch and of the
            coupling between the g-modes and the p-modes (solid curve), compared to the integration of the wave
            equation (\ref{waveeqap}), ignoring the coupling with
            p modes (red, dashed curve). The vertical,
          lines indicate the frequencies of the radial
          modes. }}
         \label{ps_full2}
          \end{figure}

Next we consider the full numerical solutions for our model~1b, located in the
core-helium-flash evolution phase. The period spacings derived from the
\adipls\ results for this model are shown in Figure~\ref{ps_full2} (solid
curve). The existence of closely-spaced pronounced dips in the period
spacing makes it harder to identify the dips associated with
the  acoustic resonant frequencies in this case.  To help with that identification,
we mark the frequencies of the radial acoustic modes (vertical lines) and recall that the dips produced by the coupling
between p and g dipole modes should be positioned roughly mid way between consecutive
radial modes. Indeed, single or double dips of greater depth than
their neighbors are found at the expected frequencies. Comparison
of the period spacing derived from the full solutions (solid curve) with that
derived considering only the g modes
(red, dashed curve) shows that the two are similar everywhere, except at the frequencies of these
more pronounced dips. This confirms that the more pronounce dips are
produced by the coupling  between the p and g modes and  excludes
that this coupling is the cause for the other dips. 
Using our analytical model for the case of having a glitch but no coupling
 (equation~(\ref{ps_glitch})) we find that the less
 pronounced dips are caused by the outer spike
 in the \brunt (Figure~\ref{figeigen}b), that is, the glitch at the
 hydrogen-burning shell. Moreover, we also confirm, based on the
 analytical model, that despite the glitch being located relatively
 far from the center of the cavity (at $\omega_{\rm g}^\star/\omega_{\rm g}= 0.075$) the larger-scale
 modulation seen in the period spacing (on a scale of about 10 glitch-induced dips) is explained by the sampling
 effect discussed in section~\ref{sec:PS_gmode}.

The separation between glitch-induced dips in model~1b is similar to
the width of the dips associated with the acoustic resonant
frequencies, which makes it difficult to interpret the combined effect
of the glitch and the mode coupling in this case. Nevertheless, the comparison between
the dips at $\approx 18.1\muHz$ and $23.2\muHz$, indicates that the
combined effect is the same as for model~1a. The latter dip is placed
at a glitch-induced hump and, consequently,  has its depth increased,
while the former dip is placed at a glitch-induced dip and has its central depth reduced,
forming a double-dip structure.  The observational impact of these double-dip
structures will be discussed in section~\ref{evolution}.

\section{Observing glitches in red giants}
\label{evolution}

Next, we search an extensive set of stellar models of various masses to locate
the stages of evolution where one could potentially observe the seismic
signature from buoyancy glitches in red giants.
Our stellar models are derived using the `default' work inlist of MESA-v5271
\citep{paxton11,paxton13} with the only change that we turned off mass
loss.  These canonical models do not include diffusion or extra mixing beyond
convection defined by the classical Schwarzschild
criterion~\citep{schwarzschild1906}. Our search comprises tracks ranging 1.0-3.0\msol, all roughly
with solar abundance, spanning the entire evolution from the bottom of
the red-giant branch to near the end of the asymptotic-giant branch. 
To check that we obtain consistent results we also derived \astec\ tracks
for 1.0\msol\ to near the tip of the red-giant branch and for 2.4\msol\ to
the end of helium-core burning.  The frequency calculations based on the full
numerical solution shown in this section were made using GYRE
\citep{townsend13}, but spot checks were made to verify that these results
were consistent with what we obtained using \adipls. 

For each model, we calculate the dipole g-mode frequencies by solving
equation~(\ref{waveeqap}) numerically, with $K$ defined by equation~(\ref{k2_gmode})
as described in  section~\ref{gnumerical}; hence, neglecting the coupling to the acoustic
cavity. The calculation is restricted to within a 
\numax/2-wide range centered around the solar-scaled \numax$\propto
g/\sqrt{T_\mathrm{eff}}$, where  $T_\mathrm{eff}$ is the effective
temperature and \numax\ is the frequency
of maximum oscillations power. This range is roughly equal to the full
width at half maximum of the excess power observed for
solar-like oscillations \citep{stello04,kjeldsen05,mosser12a}. 
From the resulting frequencies we then derive the series of pairwise period
spacings, \dP, and calculate an index of glitch-induced variation in \dP\  
to determine if the g-modes are effected by a glitch.
We tested two indices, both showing consistent results. One was 
simply the RMS of the period spacings and the other was the height of the
strongest peak in the Fourier transform of the series of period spacings versus
period.  The latter is shown for a section of the 1\msol\ track in
Figure~\ref{rgbgrid}(a) indicating a region bracketed by the vertical dotted
lines where the index is above twice the floor level. This phase is
therefore identified as showing excess variation in \dP.
Following the approach described in relation to Figure~\ref{echelle}b
(section 4), we also derive the large frequency separation for radial
modes, \dnu.

\subsection{Before helium ignition}
Along the red giant branch we find glitch-induced variations
only at one particular phase in evolution lasting roughly 5-10 million
years. Interestingly, this coincides with the luminosity bump.
\begin{figure}
\includegraphics[width=8.8cm]{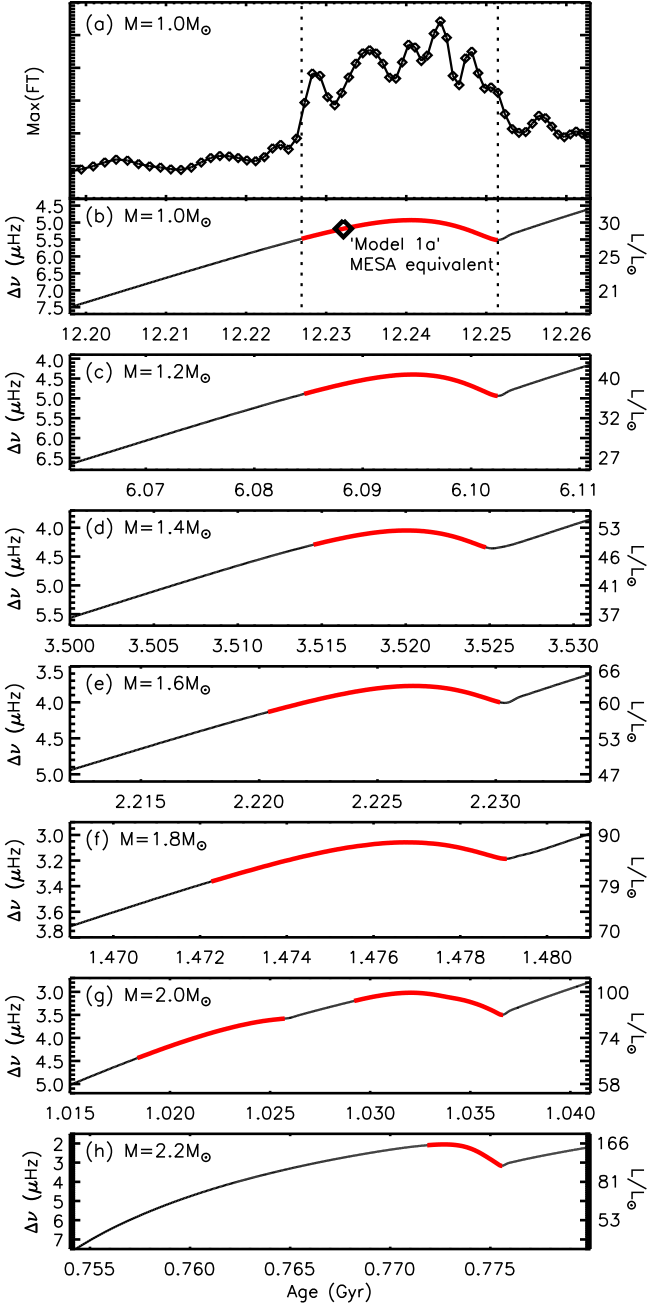}
\caption{Panel (a): Maximum signal in the
Fourier transform of the series of period spacings along the 1\msol\
track. Vertical dotted lines bracket the region of glitch-induced variation in
\dP. Panels (b-h): Close-up of the evolution near the red giant branch luminosity
bump as a function of age for models with fully or partially-degenerate
cores. Thick red curves indicate phases of excess variation in \dP. A
\mesa- equivalent of model 1a discussed in
sections~\ref{structure}-\ref{fullnumerical} is shown, in panels (a) and (b). 
\label{rgbgrid}} 
\end{figure} 
Figure~\ref{rgbgrid}(b-h) summarizes the results near the bump for  low-mass
models ($1.0 \leq M \leq 2.2\mathrm{M}_\odot$). Models beyond
$2.2\mathrm{M}_\odot$ do not show the bump because the
glitch from the first dredge-up is not reached by the hydrogen-burning
shell until after the model is past the tip of the red-giant branch.  
The thick red curves indicate the phases of excess variation in \dP.  
This excess variation can be attributed to the glitch left by the dredge-up as illustrated by model~1a (Figure~\ref{rgbgrid}b).
The only exception to this picture is along the 2.0\msol\ track, which shows an extra
slightly earlier phase of excess variation arising from a subtle but
interesting combination of effects, also resulting in the extra
luminosity bump we see for this mass at 1.026 Gyr.  
During the main-sequence phase the gradually retreating convective core
leaves a steep gradient in molecular weight (hence a spike in the buoyancy
frequency) where the convection reached its maximal extent at young age. 
For models below 1.8\msol, the gradient is smoothed away by the hydrogen-burning shell, which is later established at almost the same location.
However, for models of roughly 2.0\msol, the hydrogen-burning shell starts
at a smaller radius relative to this gradient, and the gradient therefore
survives for a while.  This allows the star to evolve 
to the point where the local wavelength becomes comparable to the scale of
the associated spike in the buoyancy frequency, giving rise to the first phase
of excess variation in \dP\ that ends when the hydrogen-burning shell
finally reaches the location of the spike.  In more massive stars that same
spike is erased by the first dredge-up before the spike appears as a glitch for
the gravity waves.

\subsection{After helium ignition}
\begin{figure}
\includegraphics[width=8.8cm]{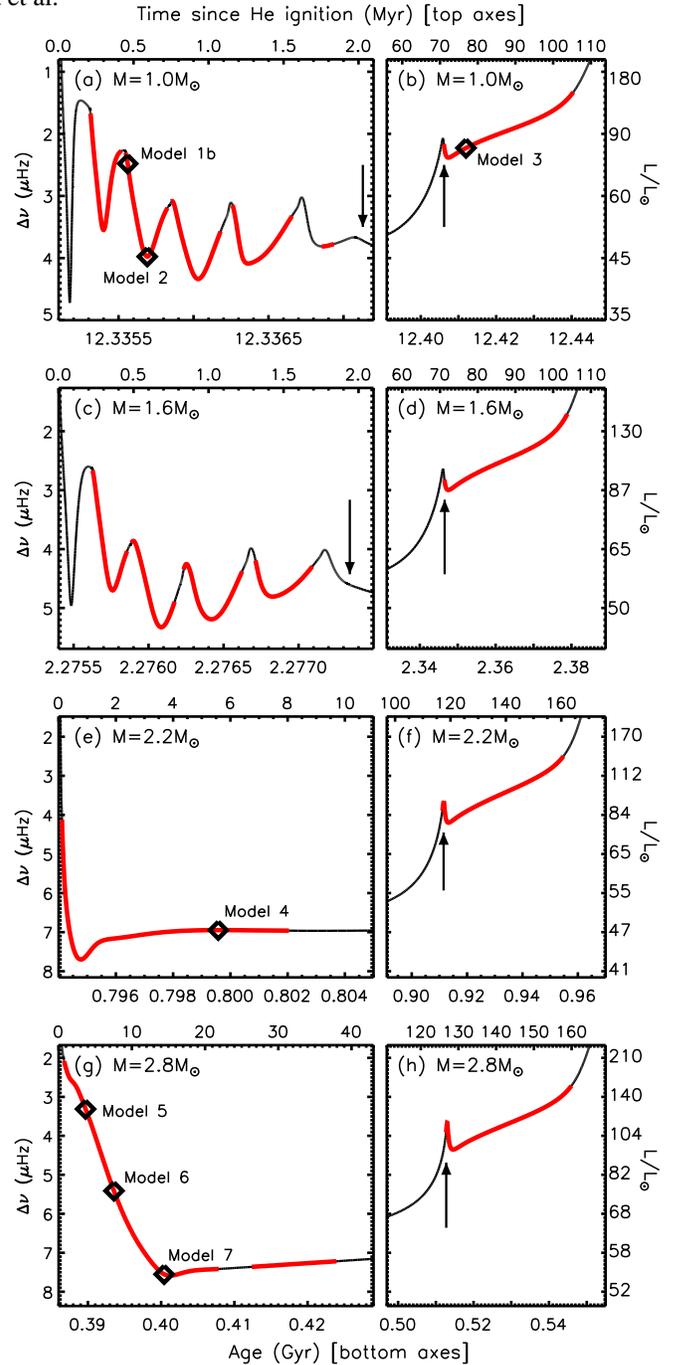}
\caption{Early (left) and late (right) stages of helium core burning. Thick
red curves indicate phases of excess variation in \dP. The time in Myrs
since helium ignition at the tip of the red giant branch is indicated along
the top axis of each panel. Down/Up-ward pointing arrows show the start/end
of quiescent helium-core burning (see text).  The left- and
  right-side annotation of the ordinate applies to both panels.
\label{hbgrid}}
\end{figure} 
In Figure~\ref{hbgrid} we show the result for post-helium ignition
tracks with masses 1.0\msol, 1.6\msol, 2.2\msol, and 2.8\msol. Again, thick
red curves indicate evolution phases showing excess variation in \dP. 
The downward-pointing arrows indicate when the last off-center helium
sub-flash and associated convection zone reaches the center, signifying the
start of quiescent helium-core burning in the models with degenerate
cores before helium ignition (Figure~\ref{hbgrid}a,c).  There is no such equivalent for the
higher-mass models, in which a more gentle at-center helium
ignition starts immediately at the tip of the red giant branch.   
The upward-pointing arrows mark the end of helium-core burning at the
so-called asymptotic-giant-branch bump, and the subsequent asymptotic-giant-branch phase. In the following we will discuss each phase in turn where we
see excess variation in \dP. 

\subsubsection{Low-mass stars} 
\begin{figure*}
\includegraphics[width=17.6cm]{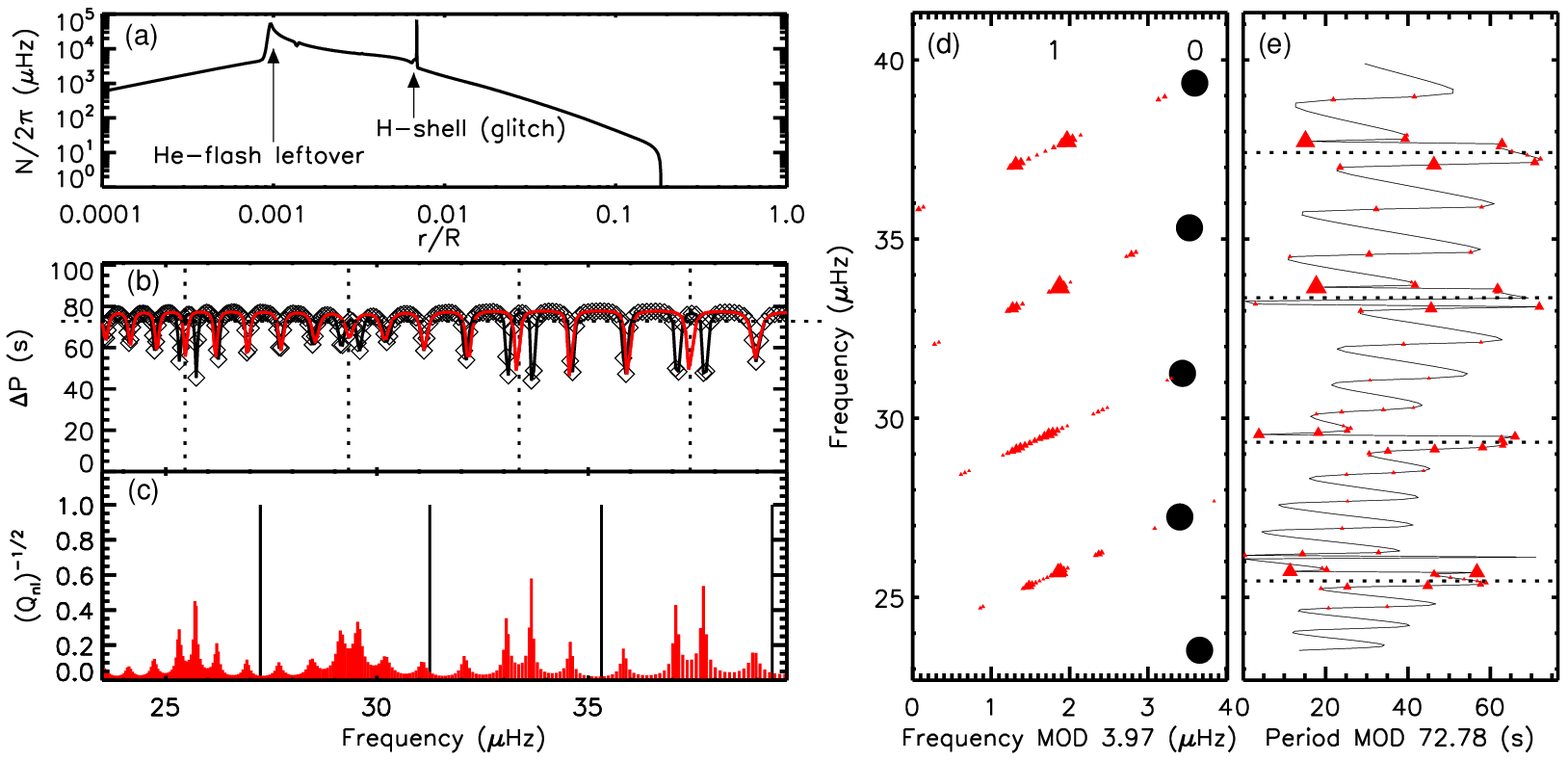}
\caption{ Model 2  ($M=1.0$~\msol\ ; \numax $ = 31\,$\muhz) - 
(a) Buoyancy frequency with key features indicated. 
(b) Period spacing of pure g modes (as in section~\ref{sec:PS_gmode}) (red) and full numerical
solution using GYRE (black). The vertical dotted lines indicate the
approximate position of the dipole acoustic modes. They have been
positioned relative to the nearest radial mode in agreement with \cite{stello14} (see also \cite{huber10,montalban10}). Horizontal dotted line marks \dPa. 
(c)  Pseudo amplitude spectrum based purely on mode inertia, normalized to the radial modes. Dipole modes are shown in red and radial modes in black. 
(d) \'Echelle diagram. The abscissa is the frequency modulo \dnu. Symbol size follows the peak heights in panel (c). 
(e) Period \'echelle diagram. Symbol sizes as in panel (d). The abscissa is the period modulo \dPa. Dotted lines indicate the
approximate position of the dipole acoustic modes. Black curve connects all dipole modes. 
\label{multipanel_M10_Mode18060}} 
\end{figure*}

Along the 1.0\msol\ track we see repeated intervals of excess variation 
during the initial helium sub-flashing phase. Each of these intervals are 
interspersed by short off-center helium burning sub-flashes where the
g-mode cavity is split in two \citep{bildsten12}. If
both g-mode cavities are taken into account, the resulting effect
on \dP\ during this cavity split differs significantly from what is
presented by \citet{bildsten12}, who ignored the inner cavity in their
analysis. The results including both cavities will be discussed in a
forthcoming paper. The intervals with only one g-mode cavity are
illustrated by model~1b, discussed in
sections~\ref{structure}-\ref{fullnumerical}, and model~2
(Figure~\ref{hbgrid}a).  In Figure~\ref{multipanel_M10_Mode18060} we show a
multi-faceted view of model~2 including its core structure and the glitch
effect on the observed frequencies. 
Model~2 is similar to
model~1b
except that it has a lower luminosity, hence
larger \numax\, and is therefore more likely to represent a case where
\dP\ can be measured in observational data \citep{mosser14,grosjean14}. 
As in model~1b, we see a series of
glitch-induced dips in \dP\ (Figure~\ref{multipanel_M10_Mode18060}b). 
The associated dips in mode inertia, or peaks in amplitude
(Figure~\ref{multipanel_M10_Mode18060}c), suggest that 
some modes would be observable even if they are far from the acoustic
resonant frequency. This decrease in the inertia arises because some, almost pure,
g-modes are trapped in the outer part of the g-mode cavity.
As a result, we see a 
split of the $l=1$ ridge in the \'{e}chelle diagram (Figure~\ref{multipanel_M10_Mode18060}d).  That split
is most evident where one of the glitch-induced dips coincides with an
acoustic resonant frequency (a coupling-induced dip), splitting the
coupling dip into two (Figure~\ref{multipanel_M10_Mode18060}b). 

\begin{figure}
\hspace{-0.5cm }
\includegraphics[width=9cm]{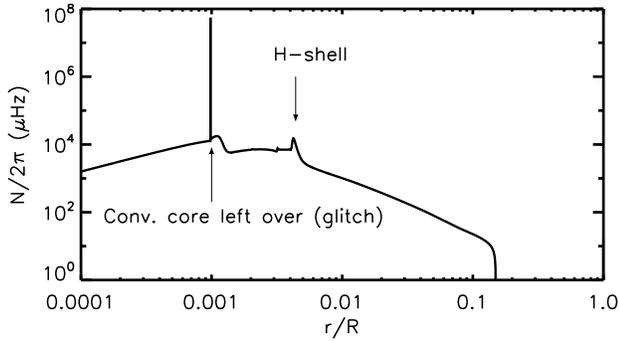}
\caption{Model 3 ($M=1.0$~\msol\ ; \numax $ = 15\,$\muhz) - Buoyancy frequency with key features indicated.
\label{multipanel_M10_Model18890}} 
\end{figure}

In the following quiescent helium-core burning phase we see no significant
variation in \dP\ for our canonical models.  
However, towards the end of core burning (Figure~\ref{hbgrid}b), the retreating
convective core leaves a sharp glitch, which results in very high-frequency
variation in \dP\ at the asymptotic-giant-branch bump and the early
helium-shell burning phase. Figure~\ref{multipanel_M10_Model18890} shows the
buoyancy frequency of model 3, which is representative for models in this phase.
The glitch is located closely to the center of the
cavity (at $\omega_{\rm g}^\star/\omega_{\rm g}\sim 0.5$). Hence, the
induced period-spacing variations occur over a scale comparable to the
separation between two consecutive modes, which results in a low-frequency
modulation in \dP\ on top of the high-frequency variation, as discussed in
section~\ref{sec:PS_gmode} and illustrated by the analytical result in
Figure~\ref{PS_gmode}c. The numerical result including adiabatic frequencies and inertias of such
models is still under investigation and will be presented in a
forthcoming paper.
The \dP\ variations indicated at the early helium-shell burning phase
(Figure~\ref{hbgrid}b, d, f, h) are
similar for all masses that we investigated, and originate from the same
physical reasons as discussed above for model~3. 
Moreover, all models that ignite helium in a degenerate core, which include
the models shown in Figure~\ref{hbgrid}c, 
show quite similar behavior to the 1.0\msol\ case, and will not be discussed
further.  

\subsubsection{High-mass stars}
\begin{figure*}
\includegraphics[width=17.6cm]{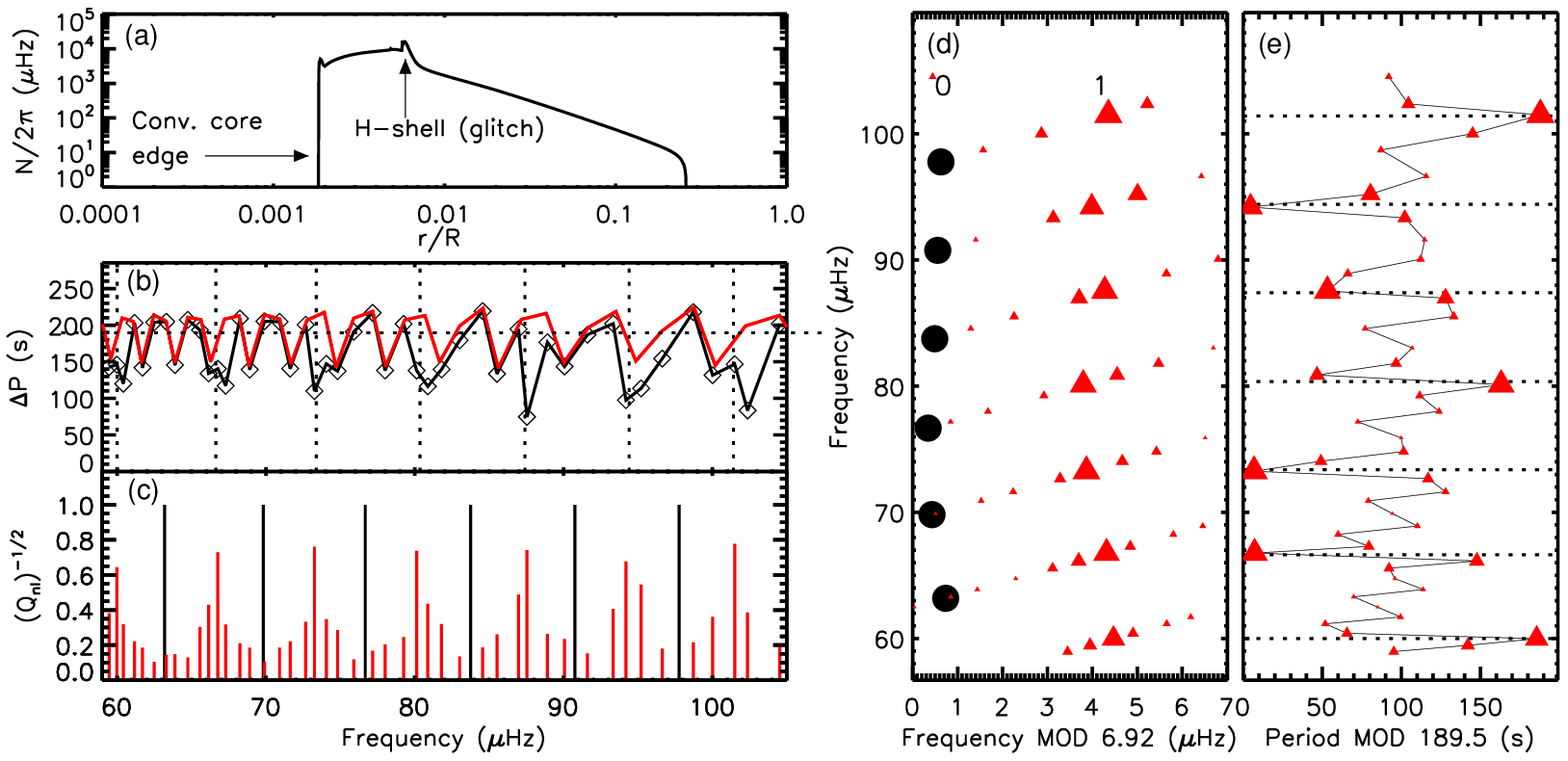}
\caption{Model 4 ($M=2.2$~\msol\ ; \numax $ = 84\,$\muhz) - notation as in Figure~\ref{multipanel_M10_Mode18060}.
\label{multipanel_M22_Model5335}} 
\end{figure*}

Moving on to a case where helium ignites in a partially degenerate
core, we see excess variation in the early stages of helium core burning as
illustrated along the 2.2\msol\ track (Figure~\ref{hbgrid}e). This variation
originates from the glitch at the hydrogen-burning shell. We show a
representative model in Figure~\ref{multipanel_M22_Model5335}. 

\begin{figure*}
\includegraphics[width=17.6cm]{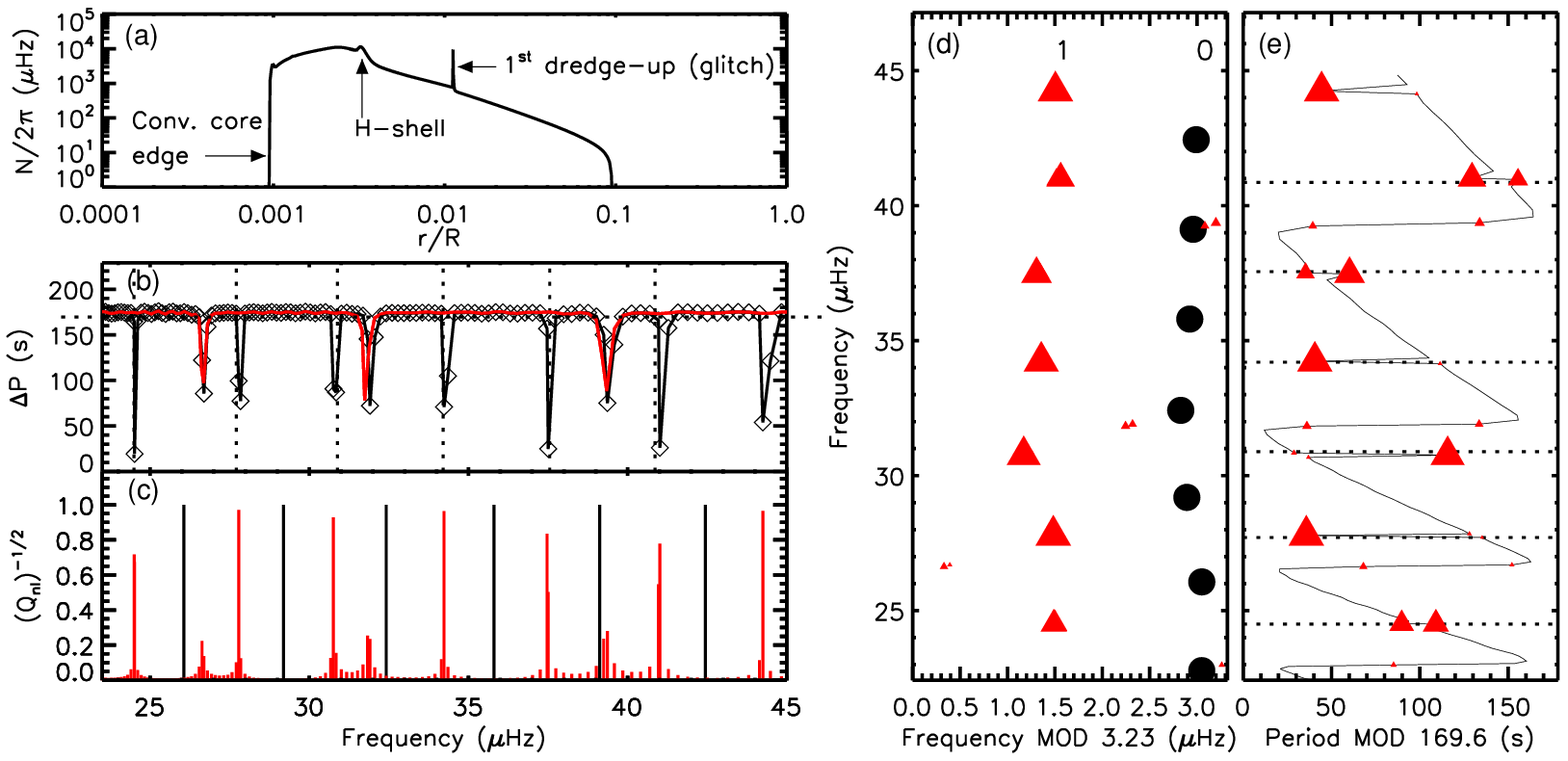}
\caption{Model 5 ($M=2.8$~\msol\ ; \numax $ = 34\,$\muhz) - 
notation as in Figure~\ref{multipanel_M10_Mode18060}.
\label{multipanel_M28_Model716}} 
\end{figure*}

\begin{figure*}
\includegraphics[width=17.6cm]{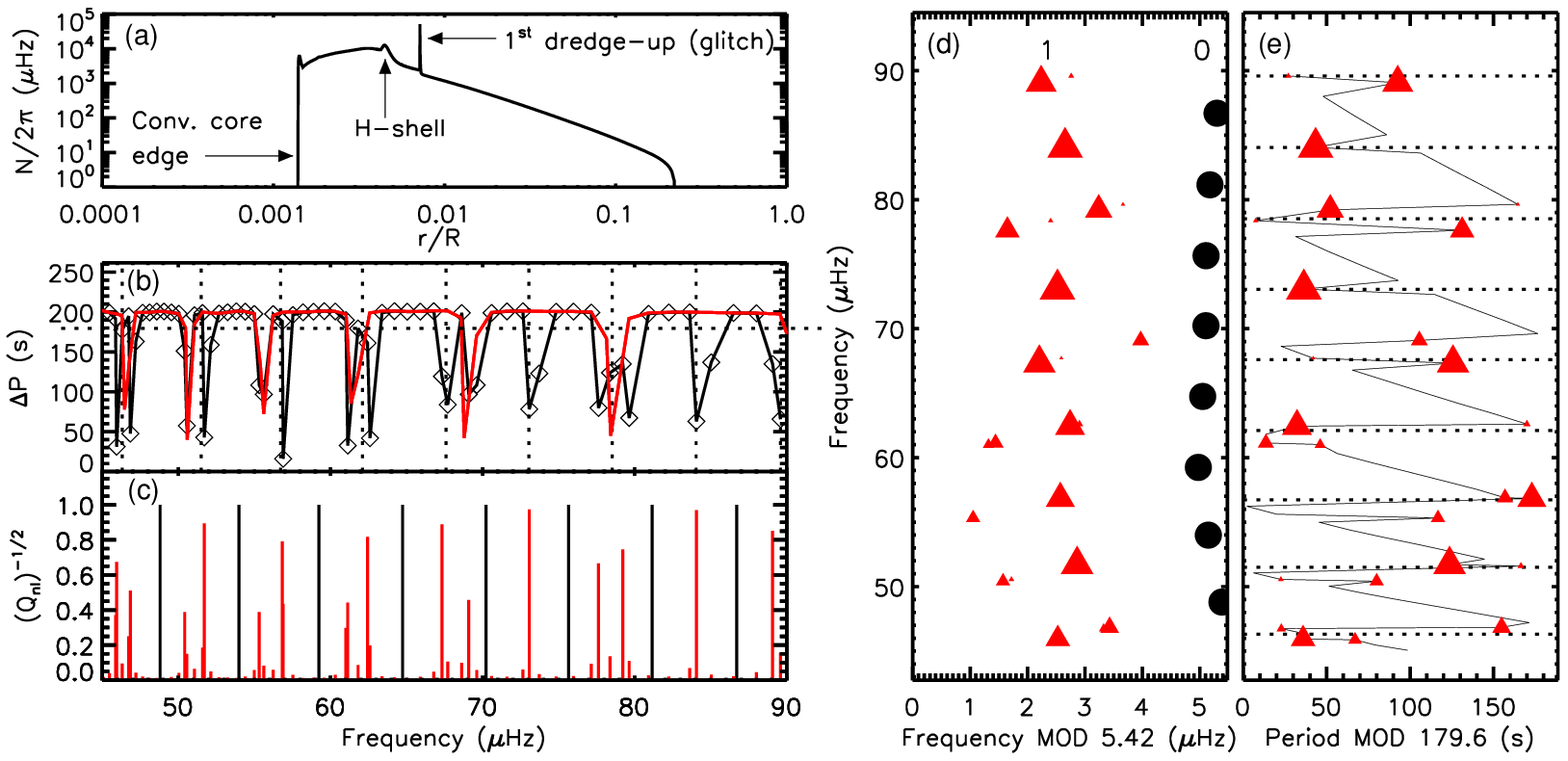}
\caption{Model 6 ($M=2.8$~\msol\ ; \numax $ = 65\,$\muhz) - 
notation as in Figure~\ref{multipanel_M10_Mode18060}.
\label{multipanel_M28_Model790}} 
\end{figure*}

\begin{figure*}
\includegraphics[width=17.6cm]{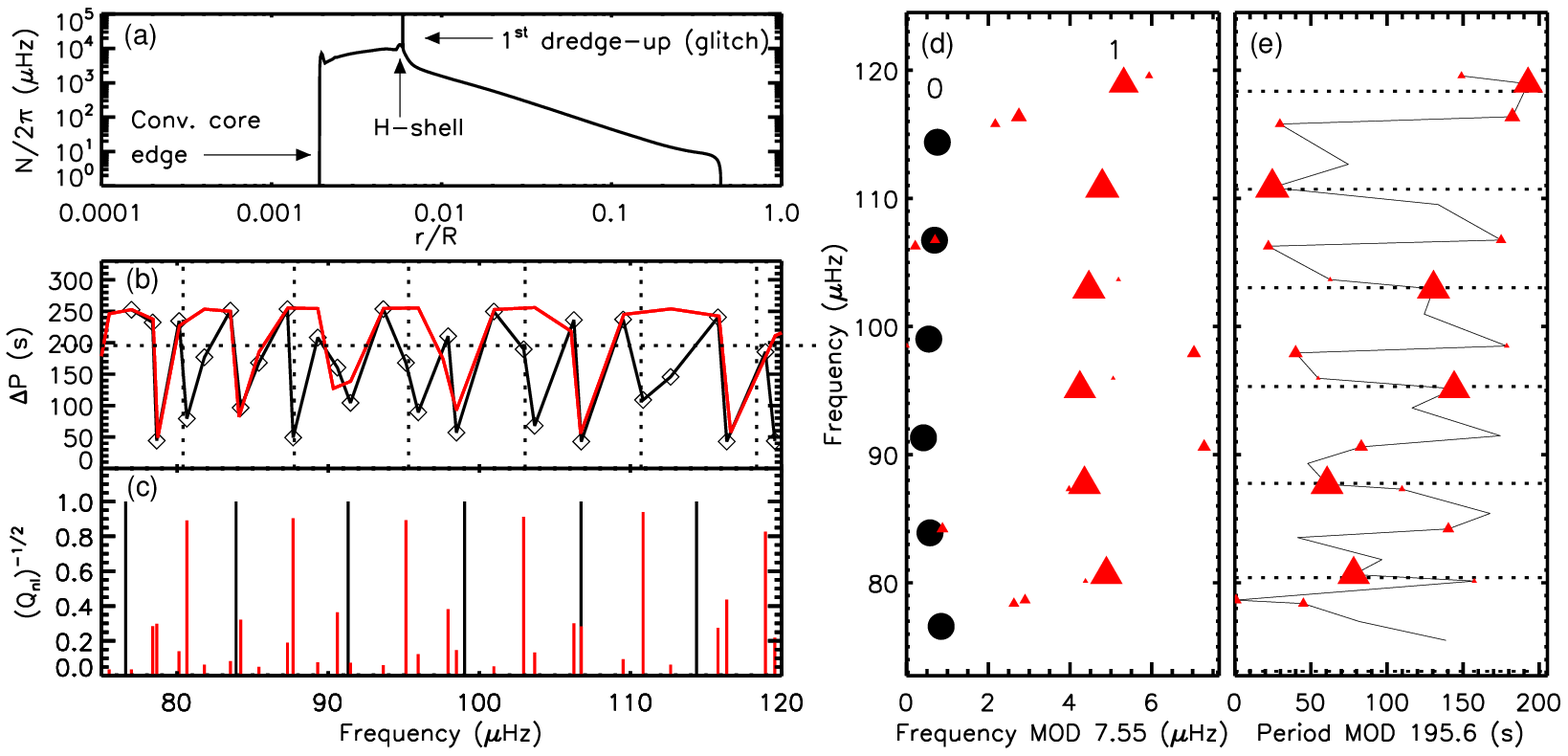}
\caption{Model 7 ($M=2.8$~\msol\ ; \numax $ = 100\,$\muhz) - 
notation as in Figure~\ref{multipanel_M10_Mode18060}.
\label{multipanel_M28_Model853}} 
\end{figure*}

Finally, representative of stars igniting helium in a non-degenerate core,
the 2.8\msol\ track shows two phases of excess variation during  
early stages of helium-core burning (Figure~\ref{hbgrid}g).  The
first phase is the `high-mass' non-degenerate-core equivalent to what
we saw in the low-mass degenerate-core models near the red-giant-branch
bump, where the glitch from the first dredge-up `enters' the g-mode cavity
(Figures~\ref{rgbgrid} and \ref{Brunt_both}c). As in the red-giant-branch bump cases, the
\dP\ variation vanishes when the hydrogen-burning shell reaches and
smooths out the glitch,
but here this occurs after the model has become a quiescent helium-core burning
clump star with \dnu\ $\sim\,7.5\,$\muhz.
Figures~\ref{multipanel_M28_Model716}, \ref{multipanel_M28_Model790}, and \ref{multipanel_M28_Model853}
show three examples along this phase of evolution. 
Like in model~2, the glitch in these three models is expected to cause
relative high amplitudes in the frequency spectrum for the almost pure g
modes located at dips in \dP\ (panels b and c). The \'echelle diagram can
therefore appear to show a dominant spacing between strong modes that is
significantly larger than the underlying period spacing between adjacent
modes (Figure~\ref{multipanel_M28_Model790}d). 
A second phase of variation occurs due to a glitch 
that was built up near the edge of the convective core during helium ignition and
its subsequent maximal extent. However, we do not show an
example of this phase because, in our
models, the variation only shows up with a relatively low amplitude (\dP\ $\sim 10\,$
sec) and dips in \dP\ that are rather broad and widely separated,
making it very difficult to detect when the coupling to the acoustic modes
is included.

\subsection{Discussion} 
Due to the glitch-induced variation in \dP\ around \dPa,
one can choose to use either \dPa\ (horizontal dotted line in panel b) or the maximum
period spacing to generate the period \'echelle (see for example
Figure~\ref{multipanel_M28_Model790}b).  We chose to use \dPa, which
in some cases creates one overall `S' shape per radial mode order as in the
glitch-free case (see 
high-frequency end of Figure~\ref{echelle}; see also Fig.1c of
\cite{bedding11}), but modulated with the glitch-induced variation on top
as in Figure~\ref{multipanel_M10_Mode18060}e and
~\ref{multipanel_M22_Model5335}e.  Had we chosen to use  
the maximum value of \dP, we would obtain one `S' shape for every
glitch-induced dip in \dP. In other cases using \dPa\ makes the period
\'echelle look very complicated with no clear pattern, such as in
Figures~\ref{multipanel_M28_Model716}e,
while the maximum value appears to create a better aligned \'echelle,
by straightening the zig-zag pattern.  The 
latter might therefore be misinterpreted as the asymptotic period spacing
of a glitch-free star.  
This could potentially explain some of the more massive stars with observed  \dPa\ reported to fall significantly outside the main ensemble in the \dPa-\dnu\
diagram by \citet{mosser14}, if indeed real stars share the frequency
behavior shown by our models.

In  search for excess variation in \dP, as summarized in
Figures~\ref{rgbgrid} and \ref{hbgrid}, 
we deliberately ignored the coupling with the envelope (p-)modes to simplify
and speedup the process. In real data, one would have to 
separate the variation in \dP\ caused by the 
coupling from the effect induced by the buoyancy glitches.  
We verified that our results presented here are
consistent with what we obtain if the p-mode coupling is included, which we
did by first fitting and removing the coupling pattern from the mixed-mode
frequencies derived from the full numerical solution (e.g. \adipls/GYRE), 
and subsequently deriving the RMS of the residual period-spacing variation. 
Fitting and removing the coupling pattern was performed along 1.0\msol\ and
2.4\msol\ tracks using the toy model for the coupling presented by
\citet{Stello12}, but could as well be done using equation~(\ref{ps_coupling}) 
\citep[see also equation~(9) in][]{mosser12b}. The analysis of glitch-induced
variations in \dP\ from real data will be presented in Stello et al. (in preparation).

Although we verified that the results summarized in Figures~\ref{rgbgrid}
and \ref{hbgrid} are similar when based on \astec, 
our \astec\ models generally showed less high-frequency variation in \dP\ 
due to the glitches being smoother, arising from numerical
diffusion in \astec, as described in section~\ref{structure}. It is also expected that including additional mixing processes
could affect the buoyancy frequency significantly, and hence alter the
signature in the frequencies, which would potentially provide a way to
test various prescriptions of mixing (Constantino et al., in preparation). 

\section{Conclusions}
\label{conclusions}
We have shown that structural glitches in the cores of red giants can
significantly affect the adiabatic properties of their mixed modes --
both mode inertias and frequencies. The modulation in mode inertia can have
strong consequences for which modes are observable. Moreover,  the
change in the frequency pattern shows up as a variation in the underlying period spacing
of pure g-modes around the fixed asymptotic value of the glitch-free case. 
Hence, assuming the period spacing follows the simple glitch-free
asymptotic behavior (equation~(\ref{ps_coupling}), see also equation~(9) in \cite{mosser12b}), can hamper
the estimate of the asymptotic (glitch-free) period spacing, \dPa . This
might explain some of the stars observed to show a period spacing that does
not follow the main ensemble of stars both along the red-giant branch and
the red clump \citep{mosser14}. 

We provide an approximate analytical solution to the wave equation in the presence of
both a structural glitch and the coupling between p and g modes. We find
that the combined effect of a glitch and mode coupling is not merely the sum of
the two. The combined effect is a modulation of
the depth of the dips at the acoustic resonant frequencies and, in
some cases, the split of these in two. The glitch-induced variations in the period spacing are
equally spaced in period, and reflect the depth at which the glitch is
located, while the amplitude of the variation is a measure of the effective
strength of the glitch. 

From an extensive set of evolution tracks of varying mass we find glitch-induced variation at the
red-giant-branch luminosity bump, at the early phases of helium-core burning,
and at the asymptotic-giant-branch bump, which signifies the beginning of
helium shell burning. We note that some of these evolution stages last for
a relatively short period of time, making the detection of glitches in such
stars a strong indicator of relative age. 

\newpage

\section*{Appendix A}
\label{apA}

In this appendix we derive the explicit form of the coupling phase,
$\varphi$, that appears in
equation~(\ref{eigen_coupling}). This phase is
uniquely  determined by the coefficients entering the solution of the wave equation in the evanescent
region $r_2\ll r \ll r_3$.  In principle these can be determined by matching the solution in the
evanescent region to that in the p-mode cavity and, subsequently,
applying an appropriate boundary condition at the photosphere.
However, in red-giant models such as those under study, we find that
$K^2$ defined by equation~(\ref{k2}) goes to $-\infty$ at some critical radius $r_{\rm c}$ located in
the evanescent region between the two cavities. The analysis of the
wave equation across this singularity is rather cumbersome and will be
considered in a separate paper (Cunha et al., in preparation). Here, 
we use, instead, the eigenvalue condition presented by
\cite{shibahashi79}, which accounts for mode coupling but not for rapid
variations in the structure (hence no glitch). Their eigenvalue condition is derived through
the asymptotic analysis of the pulsation equations, under the Cowling
approximation, for two pulsation variables, one
related to the radial component of the displacement and the other
related to the Eulerian pressure perturbation. The simultaneous use of the two equations allows the
author to avoid having to match the solutions across critical points
similar to that referred above. The result is the eigenvalue condition
\cite[][ equation~(31)]{shibahashi79}
\begin{eqnarray}
\cot\left(\int_{r_1}^{r_2}\kappa_0\rmd
  r\right)\tan\left(\int_{r_3}^{r_4}\kappa_0\rmd
  r\right) = q
\label{unno}
\end{eqnarray}
where $q$ is often called the coupling factor and is given by
\begin{equation}
q=\frac{1}{4}\exp\left({-2\int_{r_2}^{r_3}\mid\kappa_0\mid\rmd r}\right),
\end{equation}
where, as before, we used the subscript {\small 0} to indicate that this
condition is valid in the absence of a glitch. In the above,
$\kappa_0$  is an approximation to the radial wavenumber appearing in
the equations used by the author and is given by
\begin{equation}
\kappa_0^2 = \frac{\omega^2
  -N_0^2}{c^2}-\frac{L^2}{r^2}\left(1-\frac{N_0^2}{\omega^2}\right).
\end{equation}
Inside the g-mode cavity $\kappa_0\approx
L/r \, \sqrt{(1-N_0^2/\omega^2)}\approx K_0$.  Using this fact, we can
combine the conditions~(\ref{eigen_coupling})  and (\ref{unno})  to 
write,
\begin{equation}
\tan\left(\varphi\right)=\frac{q}{\tan\left(\int_{r_3}^{r_4}\kappa_0\rmd r\right)}.
\label{unno_2}
\end{equation}
Next, we note that the eigenvalue condition for pure p modes derived
by \cite{shibahashi79} (his equation~(26)) is
\begin{equation}
\int_{r_3}^{r_4}\kappa_{0{\rm a}}=m\pi,
\end{equation}
where $m$ is an integer and
the subscript ``a'' was added to indicate that this condition provides
what would be the eigenfrequencies of acoustic waves in the absence of
coupling.

Writing $\kappa_0 \equiv \kappa_{0{\rm a}}+ \delta\kappa$ and
taking  $\delta\kappa \approx \delta\omega /c$
(which is a good approximation throughout the p-mode cavity, except near the turning
points $r_3$ and $r_4$) we then have
\begin{eqnarray}
\tan\left(\int_{r_3}^{r_4}\kappa_0\rmd r\right) & = &
\tan\left(m\pi+\int_{r_3}^{r_4}\delta\kappa\rmd r\right) \nonumber
\\ & \approx &
\tan\left(\frac{\omega -\omega_{\rm a}}{\omega_{\rm p}}\right),
\label{den}
\end{eqnarray}
where $\omega_{\rm a}$ are the eigenvalues that would be obtain for
p modes in the absence of coupling and $\omega_{\rm
  p}^{-1}=\int_{r_3}^{r_4}c^{-1}\rmd r$.
Finally, using (\ref{den}) in equation~(\ref{unno_2}) we find
\begin{equation}
\varphi \approx {\rm atan}\left[\frac{q}{\tan\left(\displaystyle\frac{\omega -\omega_{\rm a}}{\omega_{\rm p}}\right)}\right].
\label{unno_3}
\end{equation}

\section*{Appendix B}
\label{apA}
Below we reproduce the expressions for the generalized \brunt and
critical frequency that appear in equations (5.4.8) and (5.4.9)
derived by \cite{gough93}.

In \cite{gough93}, the author expresses the equations describing linear, adiabatic
pulsations in terms of the Lagrangian pressure perturbation $\delta p$. After
performing the Cowling approximation, the resulting second order
differential equation for $\delta p$ is reduced to the standard
form by defining a new dependent variable $\Psi=~(r^3/g\rho f)
^{1/2}\delta p$, where $f$ is the f-mode discriminant given by
\begin{eqnarray}
f=\frac{\omega^2r}{g}+2+\frac{r}{H_{\rm g}}-\frac{L^2g}{\omega^2r},
\end{eqnarray}
and $H_{g}$ is the scale height for the gravitational acceleration
obtained following the general definition adopted by the author
that the scale height for a quantity $q$ is  $H_q=-\frac{\rmd r}{\rmd \ln q}$.
The wave equation resulting from this
variable  transformation is
\begin{eqnarray}
\frac{\rmdd\Psi}{\rmd r^2}+K^2\Psi=0,
\label{waveeqap_ap}
\end{eqnarray}
with the radial wavenumber $K$ defined by,
\begin{eqnarray}
K^2=\frac{\omega^2-\omega_{\rm
    c}^2}{c^2}-\frac{L^2}{r^2}\left(1-\frac{\mN^2}{\omega^2}\right).
\label{k2_ap}
\end{eqnarray}

In the above, $\mN$ is the generalized buoyancy frequency given
by
\begin{eqnarray}
\mN^2= g\left(\frac{1}{\mH}-\frac{g}{c^2}-\frac{2}{h}\right),
\label{N_ap}
\end{eqnarray}
where $h$ is the scale height for $g/r^2$ and is related to $H_g$ by $h^{-1}=H_g^{-1}+2r^{-1}$ and
$\mH$ is the scale height for $g\rho f/r^3$ and  is related to
other relevant scale heights in the analysis by
$\mH^{-1}=H_\rho^{-1}+H_f^{-1}+h^{-1}+r^{-1}$. Moreover, $\omega_c$ is
a generalization of the critical frequency given by
\begin{eqnarray}
\omega_c^2=\frac{c^2}{4\mH^2}\left(1-2\frac{\rmd\mH}{\rmd r}\right)-\frac{g}{h}
 \label{wc_ap}
\end{eqnarray}

Using the definition for the density scale height and the
equation for hydrostatic equilibrium, the \brunt defined by expression
~(\ref{bruntdef}) can be written as,
\begin{eqnarray}
N^2= g\left(\frac{1}{H_\rho}-\frac{g}{c^2}\right).
\label{N_ap_old}
\end{eqnarray}
Comparing this expression with expression~(\ref{N_ap}) we see that the
generalized \brunt  has  $\mH$ in the place of $H_\rho$ and includes
an additional term,  $-2/h$. As mentioned by \cite{gough93} (and seen from the definitions of $h$ and $\mH$), these differences
result from the geometry and self-gravity of the equilibrium state and,
consequently,  $\mN^2$ reduces to $N^2$  in the limit of a plane-parallel envelope
under constant gravitational acceleration. Comparison of the generalized
critical frequency with the one derived in that same limit \citep{gough07},
shows that the difference between the two is also solely the outcome
of geometry and self-gravity.

\acknowledgments
Funding for this work was provided by the University of Sydney (IRCA
grant) and by the ERC, under FP7/EC, through the project PIRSES-GA-2010-269194.
MSC and PPA are supported by the Funda{\c c}\~ao para a Ci\^encia e a
Tecnologia (FCT) through the Investigador FCT contracts of 
references IF/00894/2012 and IF/00863/2012 and by POPH/FSE (EC)
through FEDER funding through the program COMPETE. Funding for this
work was also provided by the FCT grant UID/FIS/04434/2013.
DS acknowledges support from the Australian Research Council.
Funding for the Stellar Astrophysics Centre is provided by The Danish
National Research Foundation (Grant DNRF106). The research is supported by
the ASTERISK project (ASTERoseismic Investigations with SONG and Kepler)
funded by the European Research Council (Grant agreement no.: 267864). RT acknowledges support from NASA award NNX14AB55G and NSF award ACI-1339600.

\bibliography{solar-like}

\end{document}